\documentclass{article}
\pdfoutput=1 
\usepackage{jcappub}
\usepackage{amssymb}
\usepackage{amsmath}
\usepackage{hyperref}
\usepackage{graphicx}
\usepackage{caption}
\usepackage{subcaption}

\usepackage{graphicx}
\usepackage{dcolumn}
\usepackage{bm}
\usepackage{amsmath}
\usepackage{comment}
\usepackage{physics}
\usepackage{outlines}
\usepackage{color}
\usepackage{graphicx}
\usepackage{tikz}
\usepackage{caption}
\usepackage{hyperref}
\usepackage{placeins}
\usepackage{subcaption}
\usepackage{float}
\usepackage{ragged2e}
\usepackage{orcidlink}

\newcommand{\Abacus}[1]{\textsc{#1}}
\newcommand{\Msun}{M_{\odot}}

\newcommand{\beq}{\begin{equation}}
\newcommand{\eeq}{\end{equation}}

\definecolor{darkred}{RGB}{175,0,0}
\definecolor{darkblue}{RGB}{14,0,185}
\definecolor{salmon}{RGB}{255,160,105}

\author[a,b]{Federico Semenzato,}
\emailAdd{federico.semenzato.1@phd.unipd.it}

\author[c,d]{J. Andrew Casey-Clyde}

\author[d]{Chiara M. F. Mingarelli,}
\emailAdd{chiara.mingarelli@yale.edu}

\author[a,b,e]{Alvise Raccanelli}
\emailAdd{alvise.raccanelli.1@unipd.it}

\author[a,b,e]{Nicola Bellomo,}
\emailAdd{nicola.bellomo@unipd.it}

\author[a,b,e]{Nicola Bartolo}
\author[a,b,e]{Daniele Bertacca}

\affiliation[a]{Dipartimento di Fisica e Astronomia ``G. Galilei'', Universit\`a degli Studi di Padova, via Marzolo 8, I-35131 Padova, Italy}
\affiliation[b]{INFN, Sezione di Padova, Via Marzolo 8, I-35131, Padova, Italy}
\affiliation[c]{Department of Physics, University of Connecticut, 96 Auditorium Road, U-3046, Storrs, CT 06269-3046, USA}
\affiliation[d]{Department of Physics, Yale University, New Haven, CT, 06520, USA}
\affiliation[e]{INAF - Osservatorio Astronomico di Padova, Vicolo dell'Osservatorio 5, I-35122 Padova, Italy}

\title{
Cross-Correlating the Universe: \\
The Gravitational Wave Background and Large-Scale Structure}

\abstract{
The nature of the gravitational wave background (GWB) is a key question in modern astrophysics and cosmology, with significant implications for understanding of the structure and evolution of the Universe. 
We demonstrate how cross-correlating large-scale structure (LSS) tracers with the GWB spatial anisotropies can extract a clear astrophysical imprint from the GWB signal. 
Focusing on the unresolved population of supermassive black hole binaries (SMBHBs) as the primary source for the GWB at nanohertz frequencies, we construct full-sky maps of galaxy distributions and characteristic strain of the GWB to explore the relationship between GWB anisotropies and the LSS. 
We find that at current pulsar timing array (PTA) sensitivities, very few loud SMBHBs act as Poisson-like noise. 
This results in anisotropies dominated by a small number of sources, making GWB maps where SMBHBs trace the LSS indistinguishable from a GWBs from a uniform distribution of SMBHBs. 
In contrast, we find that the bulk of the unresolved SMBHBs produce anisotropies which mirror the spatial distribution of galaxies, and thus trace the LSS. 
Importantly, we show that cross-correlations are required to retrieve a clear LSS imprint in the GWB. 
Specifically, we forecast the distinguishability of this LSS signature at a $3\sigma$ level in near-future PTA experiments that probe angular scales of $\ell_{\text{max}} \geq 42$, and $5\sigma$ for  $\ell_{\text{max}} \geq 72$ in optimistic settings.
These values assume that anisotropic GWB maps can be reconstructed at the corresponding angular resolution and that loud sources above a resolvability threshold can be identified and removed; actual sensitivities will depend on instrumental noise, map-making uncertainties, and source-subtraction accuracy.
Our approach opens new avenues to employ the GWB as an LSS tracer, providing unique insights into SMBHB population models and the nature of the GWB itself. 
Our results motivate further exploration of potential synergies between next-generation PTA experiments and cosmological tracers of the LSS.
}

\begin{document}

\maketitle

\section{Introduction}

All Pulsar timing array (PTA) experiments have now reported evidence for a gravitational wave background (GWB)~\cite{agazie_nanograv_2023, eptacollaboration_second_2023, reardon_search_2023, xu_searching_2023} at nanoHertz frequencies. 
The primary source of the GWB is likely the superposition of low-frequency GWs from the cosmic merger history of supermassive black hole binaries (SMBHBs;~\cite{foster_constructing_1990, sesana_stochastic_2008,Becsy_2022, agazie2024_ACC}), though there may be contributions from cosmological sources~\cite{Alba_2016, Contaldi_2017, Bartolo_2020, Valbusa_Dall_Armi_2021, Afzal_2023}. 
In fact, the very large amplitude of the GWB raises important questions about its origin, and this is under active investigation, see, e.g., ref.~\cite{satopolito2023nanogravsbigblackholes}. 

As PTA experiments mature, add pulsars and improve instrumentation, we will be able to fully characterize the GWB. 
This characterization includes GWB map-making~\cite{Mingarelli_2013, nanograv23_anisotropy}, and more precisely measuring the GWB's amplitude and strain spectrum~\cite{agazie_nanograv_2023, eptacollaboration_second_2023, agazie2024_ACC}. 
These measurements will provide more clues as to the composition of the GWB, and the physics sourcing it: a GWB sourced by SMBHBs will show signs of discreteness in the characteristic strain spectrum in the form of excursions from the $h_c(f) \propto f^{-2/3}$ power-law~\cite{phinney_practical_2001}, due to a relative over-or-under-density of SMBHBs at a specific frequency, while a cosmological background will be smooth.  

\begin{figure}[t] 
  \centerline{
  \begin{subfigure}{0.5\textwidth} 
      \includegraphics[width=\columnwidth]{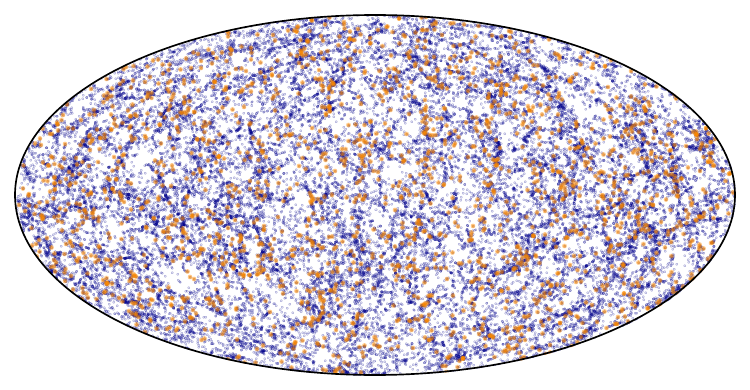}
      \caption{SMBHBs and galaxies trace the underlying LSS} 
  \end{subfigure}
  \begin{subfigure}{0.5\textwidth} 
      \includegraphics[width=\columnwidth]{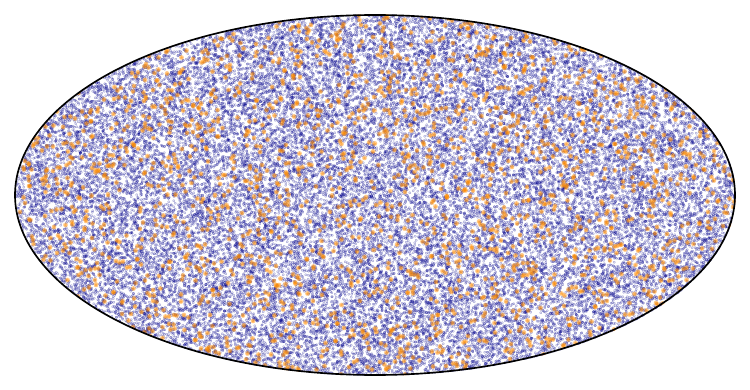}
      \caption{Uniform distribution of SMBHBs and galaxies} 
  \end{subfigure}}
  \captionsetup{width=\textwidth}
  \caption{SMBHBs are hosted by massive galaxies, which trace the LSS. 
  This figure provides an illustrative representation of the underlying physical distribution of these sources and their host galaxies, not a direct PTA observation of individual sources. 
  Here we illustrate how a LSS-informed map (a) of galaxies (blue) and SMBHBs in these galaxies (orange) differ from one where the positions of both are isotropized (b). 
  In this paper we show that LSS imprints spatial anisotropies in the GWB, even though only 1-4\% of galaxies host SMBHBs~\cite{agazie_nanograv_2023}. 
  This motivates our exploration of cross-correlations between galaxies and the GWB, whose statistical properties are accessible to PTAs. }
  \label{fig:cross_maps}
\end{figure}

These binaries are hosted at the center of massive galaxies which result from galaxy major mergers~\cite{Begelman_1980, Kormendy_1995}, and thus trace the history of galaxy evolution.
Therefore, the spatial distribution of SMBHBs is expected to be correlated to that of galaxies, see figure~\ref{fig:cross_maps}; or, said differently, a GWB generated by SMBHBs inherits the spatial distribution of their host galaxies. 
This opens the possibility of exploiting GWB anisotropies to probe the LSS, as these anisotropies act as a novel tracer of the underlying dark matter distribution, the so-called {\it cosmic web}. 
Although all GWB anisotropies can be traced back to the LSS, nearby SMBHBs can introduce strong anisotropies that hide the cosmic web imprints in the GWB maps. 
This kind of loud-source anisotropy can be searched for and modeled using current techniques developed by, e.g., ref.~\cite{Mingarelli_2013,taylor2020}. 
Targeting LSS-induced anisotropies and comparing them with galaxy clustering therefore provides a clear test of the nature of the GWB. 
Indeed, non-zero cross-correlations between the LSS and the GWB would support SMBHBs as the primary source of the GWB~\cite{Mingarelli_2017, sah2024, sah2024_imprints, lemke2024}. 

It is important to distinguish the statistical characterization of the GWB anisotropy, which relies on the cross-correlation of timing residuals from a network of pulsars to reconstruct the GWB angular power spectrum ($C_\ell^{\text{GWB}}$), from the localization of individual, bright continuous wave sources.
While the latter is a key science goal for PTAs~\citep{Arzoumanian_2020, Arzoumanian_2023, NNAOGRAV_CW}, our focus here is on the information encoded in the anisotropies of the unresolved GWB component, once sufficiently loud individual sources are identified and their contributions potentially removed from the data.
Localization of a single continuous wave source may suffer from degeneracies when relying solely on limited information like the Earth term for a single pulsar pair. 
In contrast, stochastic background analysis circumvents these degeneracies by leveraging the full dataset of pulsar auto- and cross-correlations. 
The achievable angular resolution~$\ell_{\rm max}$ for the GWB map is determined by the overall PTA characteristics, including the number and sky distribution of pulsars, timing precision, and observation duration, see, e.g., refs.~\citep{Mingarelli_2013,alihaimoud2020,rosado_expected_2015}.

Here we explore the correlation between GWB anisotropies and galaxy clustering. 
Recently, refs.~\cite{sah2024_imprints,sah2024} investigated how cross-correlations can provide unique insights into the cosmic evolution of SMBHBs by forecasting the detectability of the cross-correlation between upcoming galaxy surveys and future PTA measurements of the GWB. 
While they employ a theoretical evaluation of the signals validated on simulations, we focus on a rigorous simulation-based analysis not only to include the effect of single loud sources and discreteness of the SMBHB population, but also to investigate the significance of the cross-correlation measurement in a single universe.
We do this by simulating the GWB, using a mock galaxy catalog combined with a semi-analytical model of the SMBHB population. 
We then develop a general framework that leverages cross-correlations with LSS to assess how distinguishable GWB and SMBHB population models are from each other. 
We focus on distinguishing a GWB informed by LSS, and one generated from a uniform distribution of SMBHBs.
We explore the impact of cosmic variance for the first time in this kind of study, by comparing results from individual simulated realizations to those averaged over many simulations. 
Taking cosmic variance into account is crucial because neglecting it can lead to an error in estimating how detectable the LSS imprint is on the GWB.

A fully realistic anisotropy analysis requires modeling the response of each pulsar pair to the GWB, parametrizing the angular power distribution as a linear combination of basis functions, and inferring these parameters directly from the correlated timing residuals through a comprehensive Bayesian framework~\cite{alihaimoud2020, taylor2020, Ali_Ha_moud_2021, pol_taylor, konstandin2024impactcosmicvarianceptas, Grunthal_2024}. 
Such an end-to-end pipeline would naturally incorporate instrument-specific systematics, pulsar noise characteristics, and reconstruction uncertainties that affect the quality and fidelity of the resulting sky maps. 
However, in this work, we adopt a complementary approach: we assume that an anisotropic GWB map has already been successfully reconstructed through such methods. 
This optimistic setting enables us to examine the \textit{theoretical information content} of the cross-correlation signal, independently of the additional complications introduced by specific map-making algorithms. 

The paper is organized as follows. 
In Section~\ref{sec:tracers}, we describe the pipeline that is used to generate a mock galaxy distribution and GWBs. 
In Section~\ref{sec:analysis}, we introduce the statistical formalism that allows to quantify the nature of the GWB anisotropies, while in Section~\ref{sec:results} we present the LSS-induced anisotropies detectability analysis. 
We discuss the impact of our results in Section~\ref{sec:discussion}.
Appendices~\ref{app:galaxy_mock}, \ref{app:covariance}, \ref{app:ks_test}, and~\ref{app:gwb_bias_detectability} contain additional material that complements and supports the results presented in the main text.

%%%%%%%%%%%%%%%%%%%%%%%%%%%%%%%%%%%%%%%%%%%%%%%%%%%%%%%%%%%%%%%%%%%%%%%%%%

\section{Tracers of LSS}
\label{sec:tracers}

\begin{figure}[ht]
  \centering
  \includegraphics[width=0.7\columnwidth]{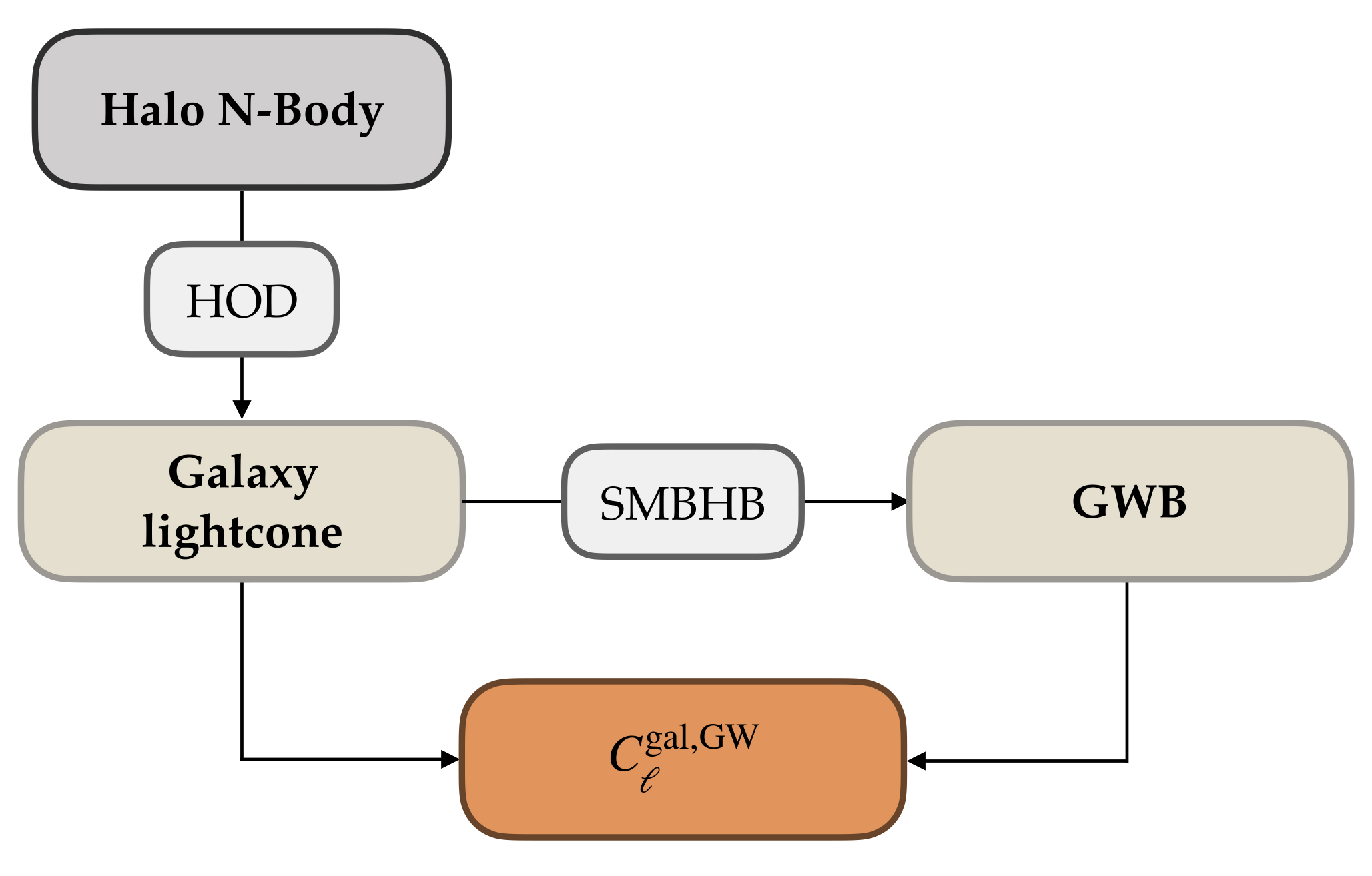}
  \captionsetup{width=\columnwidth}
  \caption{Flow-chart of our pipeline. 
  The galaxy lightcone is built from the \Abacus{AbacusSummit} suite of $N$-body simulations using the \Abacus{AbacusHOD} prescription. 
  The SMBHB populations are created using a semi-analytical model from~\cite{Casey_Clyde_2022, agazie2024_ACC}, from which the GWB is computed.
  The galaxy distribution and the GWB are then cross-correlated to study the LSS imprint on the GWB.}  
  \label{fig:flow_chart}
\end{figure}

Anisotropies in the spatial distribution of LSS tracers, such as galaxies and SMBHBs, encode information about their common evolution history.
Here we consider the anisotropic distribution of galaxy number counts and the GWs emitted by SMBHBs.
Briefly, we take a simulation-based approach, summarized in figure~\ref{fig:flow_chart}. 
Starting from the~\Abacus{AbacusSummit} suite of~$N$-body simulations, we construct mock lightcones of galaxies by populating dark matter halos using the~\Abacus{AbacusHOD} parametrization. 
We then populate these galaxies with SMBHBs using semi-analytical models from refs.~\cite{Casey_Clyde_2022, agazie2024_ACC}.
Finally, from these galaxy and SMBHB catalogs, we build galaxy number count and GWB full-sky maps and study their angular statistics.

%%%%%%%%%%%%%%%%%%%%%%%%%%%%%%%%%%%%%%%%%%%%%%%%%%%%%%%%%%%%%%%%%%%%%%%%%%

\subsection{Galaxy distribution}
\label{subsec:galaxy_distribution}

To generate full-sky galaxy maps, we employ the \Abacus{AbacusSummit} suite of $N$-body simulations~\cite{abacus_lc}. 
We use dark matter halo merger trees to construct the cosmic evolution across the radial direction, effectively building a halo lightcone.
Specifically, we use one of the two~\textit{huge} simulations (box size of~$7600 \ h^{-1}\mathrm{Mpc}$ and~$8640^3$ particles of mass~ $M_{\rm part}=2.1 \times 5 \times 10^{10} \ h^{-1}M_\odot$).
Observers are placed at the center of the box and can observe galaxies up to the half-distance of the box,~$3.75\ h^{-1}{\rm Gpc}$.

The halo lightcone catalog is then populated with galaxies using the halo occupation distribution (HOD) model~\cite{2000Peacock, 2001Scoccimarro, 2005Zheng}, which statistically connects the distribution of galaxies to their host dark matter haloes. 
We also include parameters to account for the dependence of each galaxy on its host halo mass, redshift, type, and environment, to better match current observations of galaxy clustering~\cite{white_rees_1978, Blumenthal_1984}.
We employ the~\Abacus{AbacusHOD} prescription~\cite{HOD_Yuan:2021izi} to populate the haloes with luminous red galaxies (LRGs; see appendix~\ref{app:galaxy_mock} for more details).
We validate the statistical properties of the simulated lightcone against the theoretical prediction of \textsc{CLASS}~\cite{lesgourgues2011cosmiclinearanisotropysolving}, see also appendix~\ref{app:class_validation} for additional details.
For now we focus on the density term of the galaxy number count, and in the future we will include projection effects and redshift-space distortions. 

To assign stellar masses to galaxies, we use the stellar mass ($M_*$) to halo mass ($M_h$) ratio from the \textsc{UniverseMachine} framework~\cite{universe_machine_Behroozi:2019kql}.
The~\textsc{UniverseMachine} approach parameterizes the galaxy star formation rate as a function of the halo potential well depth, redshift, and assembly history.
The redshift-dependent stellar-mass-to-halo-mass ratio fits are provided for both central and satellite galaxies.
Due to simulation bounds, the distribution is truncated outside the range ${z\in[0.1,2]}$.
This limitation is particularly relevant for low redshifts, where the highest-strain GW sources are expected to be located. 
The exploration of the low-redshift regime is left for future work. 
Here we focus on a redshift range in which LSS anisotropies are expected to be significant, assuming that low-redshift sources will be detected as continuous GW signals.

%%%%%%%%%%%%%%%%%%%%%%%%%%%%%%%%%%%%%%%%%%%%%%%%%%%%%%%%%%%%%%%%%%%%%%%%%%

\subsection{The Gravitational wave background}
\label{subsec:GWB}

Galaxies are populated by SMBHBs following a semi-analytical major-merger model calibrated to PTA measurements~\cite{chen_constraining_2019, Casey_Clyde_2022, agazie2024_ACC, 2024arXiv240519406C}.
In practice, the model combines a time-dependent Schechter stellar-mass function, a fractional pairing-rate function, a merger-timescale model, and empirical $M_*\!-\!M_{\rm bulge}\!-\!M_{\rm BH}$ relations to construct the SMBHB merger rate~$\dot\phi_{\rm BHB}$, from which different realizations of sources are sampled. 
The parameter space is constrained by PTA data, so that the procedure always returns a population whose GWB is consistent with NANOGrav measurements.
We generate $N_r=1,000$ SMBHB population realizations for the same galaxy distribution to account for cosmic variance. 
Each realization has a SMBHB population sampled from the differential number of SMBHBs per unit of total mass~$M$, mass ratio~$q$, redshift~$z$, and frequency~$f_\mathrm{GW}$, defined as~\cite{peters_gravitational_1963, hogg_distance_1999, chen_constraining_2019}:
\begin{equation}
\label{eq:diff_num}
    \mathcal{N}_{\rm BHB}(M, q, z, f_\mathrm{GW}) = \frac{d^{3} \Phi_{\rm BHB}}{d M \,dq \,dt_\mathrm{r}} \frac{dV}{dz} \frac{dt_{\rm r}}{df_{\rm r}} \frac{df_{\rm r}}{df_{\mathrm{GW}}},
\end{equation}
where~$\Phi_{\rm BHB}$ is the comoving number density of SMBHB mergers, $dV/dz$ is the comoving volume element~\cite{hogg_distance_1999}, $f_{\rm r} = f_\mathrm{GW} (1 + z)$ is the rest frame GW frequency, and~$t_{\rm r}$ is the residence timescale~\cite{peters_gravitational_1963}.
The sky- and polarization-averaged strain emitted by each SMBHB is given by~\cite{thorne_gravitational_1987, sesana_stochastic_2008, rosado_expected_2015}
\begin{equation}
    h = \frac{8}{\sqrt{10}} \frac{\mathcal{M}_{\mathrm{o}}^{5 / 3} (\pi f_{\mathrm{GW}})^{2 / 3}}{D_{L}(z)}\,,
\end{equation}
where $D_{L}(z)$ is the luminosity distance, $\mathcal{M}$ and~$\mathcal{M}_{\mathrm{o}} = \mathcal{M} (1 + z)$ are the chirp masses in the source and observer frame, respectively.
The characteristic strain of the GWB is typically parametrized as~$h_{c}(f_{\mathrm{GW}}) = A_{\rm yr} \left(f_{\mathrm{GW}} / f_{\rm yr} \right)^{\alpha}$, where~$A_{\rm yr}$ is the GWB amplitude at a reference frequency~$f_{\rm yr} = 1~\rm{yr^{-1}}$ and~$\alpha = - 2 / 3$ is the spectral index for a GWB sourced by SMBHBs~\cite{phinney_practical_2001, sesana_stochastic_2008}.
Each GWB realization is consistent with the NANOGrav results,~$A_{\mathrm{yr}} = 2.4^{+0.7}_{-0.6} \times 10^{-15}$~\cite{agazie_nanograv_2023, Casey_Clyde_2022}.

\begin{figure}[ht]
  \centerline{
  \includegraphics[width=0.75\columnwidth]{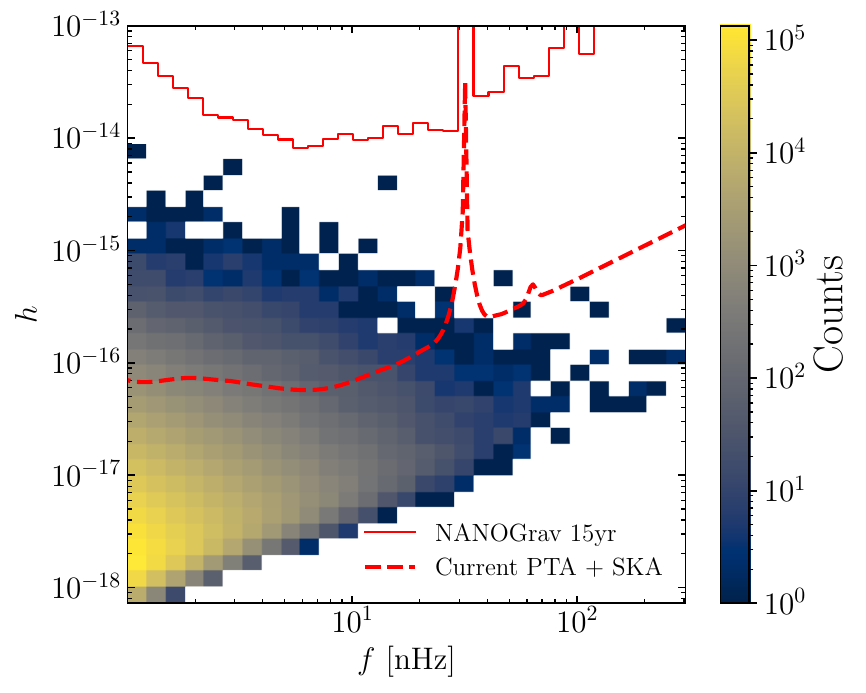}}
  \captionsetup{width=\columnwidth}
  \caption{Loud GW sources are present across all frequencies. 
  While few in number, they contribute significantly to the GWB anisotropy. 
  They must therefore be removed in order to find the LSS imprint in GWB. 
  Here we show the frequency and strain distribution for one realization of the SMBHB population, with the detection curves for current and future PTA experiments from~\cite{Xin_2021, NNAOGRAV_CW}.} 
\label{fig:smbhb_hists}
\end{figure}

In Figure~\ref{fig:smbhb_hists} we see the typical strain and frequency distribution of realization of a SMBHB population, along with the sky-averaged sensitivity curve for resolvable SMBHBs for the NANOGrav 15-yr dataset~\cite{NNAOGRAV_CW, nanograv_detector} and for future PTA experiments~\cite{Xin_2021}.
The NANOGrav 15-yr detection curve is based on the noise parameters in ref.~\cite{NG15yr_data} and on the upper limits on the GW amplitude from individual binaries from~\cite{NNAOGRAV_CW}.
The SKA sensitivity curve is computed as in~\cite{Xin_2021}, using the open-source Python package~\texttt{hasasia} and the signal-to-noise estimates from~\citep{HazbounPRD:2019}.
We set a signal-to-noise of~$\mathrm{S/N}=3$ on the GW strain, as typically assumed for detection of single sources in the nHz band.
We draw PTA-suitable pulsars from the planned SKA MID and LOW survey distributions, assuming that~$15\%$ of the detected pulsars will be suitable for PTA studies~\citep{KeaneEtAl:2015}. 
We draw pulsars with~${\rm RMS}=300\pm 100$~ns, and cadences based on those in~\cite{NG15yr_data, Xin_2021}.
We include current NANOGrav pulsars, assuming that SKA PTA will be based on extended PTA datasets.
In total, SKA1 will then
have~$\sim 675$ millisecond pulsars, and SKA2~$\sim 795$~\cite{Xin_2021}, while more optimistic projections consider even~$10^3$ pulsars~\cite{janssen2014gravitationalwaveastronomyska}.

It should be noted that these sensitivity curves do not explicitly incorporate the contribution of the GWB to the noise budget. 
At lower frequencies ($\sim$11 nHz), the GWB may decrease sensitivity by a factor of approximately 3~\citep{Xin_2021}. 
A more precise assessment of individual source detectability — especially for future experiments such as SKAO — would benefit from an iterative approach that accounts for both instrumental noise and the residual GWB~\citep{rosado_expected_2015}. 
However, the GWB is subdominant~\cite{agazie2024_ACC} at higher frequencies relevant to our SMBHB candidates, making our current method sufficient for order-of-magnitude forecasts.

Detecting LSS-induced anisotropies in the GWB will require improved angular resolution beyond current PTA capabilities. 
The SKAO, whose pathfinder MeerKAT has already independently confirmed the GWB~\citep{Miles_2024}, offers the most viable path forward. 
Although the number of measurable modes scales with the number of pulsar pairs ($N_{\rm pairs}$)~\cite{alihaimoud2020}, reconstruction at high~$\ell$ is limited by anisotropic pulsar distributions, heterogeneous timing precision, and source confusion. 
Nonetheless, the SKAO’s wider, more isotropic pulsar population and targeted selection of low-red-noise pulsars will enhance sensitivity to anisotropic structure.

%%%%%%%%%%%%%%%%%%%%%%%%%%%%%%%%%%%%%%%%%%%%%%%%%%%%%%%%%%%%%%%%%%%%%%%%%%

\subsection{Map-making and analysis}
\label{subsec:mapmaking}

We build full sky maps of our tracers~$X=\{\mathrm{gal},\mathrm{GW}\}$ (where~$X$ is either the number count for galaxies or the characteristic strain for SMBHBs) and study the spatial anisotropies of their overdensity field, defined as~\cite{peebles}:
\begin{equation}
\label{eq:od}
    \delta^X(\hat{n}) = \frac{X(\hat{\mathbf{n}})}{\bar{X}}-1\,,
\end{equation}
where $\bar{X}$ is the mean value of the field and $\hat{\mathbf{n}}$ is the line-of-sight direction.
For galaxies, we construct galaxy number count maps~$N(z_i,\hat{\mathbf{n}})$. 
For each redshift bin~$z_i$, we count galaxies within each HEALPix-distributed pixel of a skymap~\cite{Gorski_2005, Zonca2019}. 
For the GWB, we construct a map of the characteristic strain as~\cite{rosado_expected_2015}:
\begin{equation}
    h_{c}^{2}(f_\mathrm{GW},\hat{\mathbf{n}}) = \frac{\sum_{k} h_{k}^{2} f_{\mathrm{GW},k}}{\Delta f_\mathrm{GW}} \,,
\label{eq:hc2}
\end{equation}
where~$k$ runs over all sources in a given frequency bin and direction, $\Delta f_{\mathrm{GW}} = T^{-1}_{\rm obs}$ is the frequency sampling interval, and~$T_{\rm obs}$ is the total observation time.
Information on clustering properties is enclosed in the $n$-point functions of the overdensity field.
Since we are considering full-sky maps, it is more convenient to expand the overdensity field in spherical harmonics as
\begin{equation}
    \delta^X(\hat{\mathbf{n}}) = \sum_{\ell m} a_{\ell m}^X Y_{\ell m}(\hat{\mathbf{n}})\,,
\end{equation}
where~$a^X_{\ell m}$ are the spherical harmonics coefficients and~$Y_{\ell m}$ are spherical harmonics.
The observed harmonic coefficients can be decomposed into a signal~$s_{\ell m}^X$ and noise~$n^X_{\ell m}$ component, so that~${a_{\ell m}^X=s_{\ell m}^X+ n^X_{\ell m}}$.
The total \textit{theoretical} two-point function in harmonic space, i.e., the angular power spectrum~$\tilde{C}_\ell$, for any two tracers is then:
\begin{equation}
    \langle a^X_{\ell m} a^{Y*}_{\ell' m'} \rangle = \delta^K_{\ell\ell'} \delta^K_{mm'} \tilde{C}^{XY}_\ell,
\end{equation}
where~$\delta^K$ is the Kronecker delta and $\langle \cdot\rangle$ denotes the ensemble average.
The total theoretical angular power spectrum can be written as 
\begin{equation}
    \tilde{C}^{XY}_\ell = C^{XY}_\ell +  N^{XY}_\ell,
\end{equation}
where~$C^{XY}_\ell$ is the clustering signal arising from the correlated spatial distribution of tracers~$X$ and~$Y$ we aim to constrain, either~$C_\ell^{\rm gal,gal}$~\cite{peebles, landy_1993, hamilton_1993_acf}, $C_\ell^{\rm GW,GW}$~\cite{Mingarelli_2013, Cusin_2017, Cusin_2018, Cusin_2019, Bartolo_2019} or the cross term~$C_\ell^\mathrm{gal,GW}$, and~$N^{XY}_\ell$ is the noise power spectrum.
For auto-correlations~(\mbox{$X=Y$}),~$N^{XX}_\ell$ is the standard Poisson shot noise term. 
For cross-correlations~(\mbox{$X \neq Y$}), if the tracers are perfectly independent samplings of the underlying field, $N^{XY}_\ell = 0$. 
However, if one population of tracers is a subset of or directly related to the other (e.g., SMBHBs hosted in galaxies), a non-zero correlated shot noise term~$N^{XY}_\ell$ can still exist \citep{cusin2025measuringanisotropiesptaband}, albeit being totally negligible in the case at hand. 
In our simulation-based approach, these effects are naturally included in the \textit{observed} angular power spectrum~$\hat{C}_\ell^{XY}$, defined as
\begin{equation}
    \hat{C}^{XY}_\ell 
    = \frac{1}{2\ell+1} \sum_{m=-\ell}^{\ell} a_{\ell m}^{X} \left( a_{\ell m}^{Y} \right)^*\,.
\end{equation}
Since for each multipole~$\ell$ the number of~$a_{\ell m}$ modes is finite, its estimation is limited by cosmic variance.

%%%%%%%%%%%%%%%%%%%%%%%%%%%%%%%%%%%%%%%%%%%%%%%%%%%%%%%%%%%%%%%%%%%%%%%%%%

\section{Distinguishing anisotropy models}
\label{sec:analysis}

Previous cross-correlation studies have addressed the issue of distinguishing between different underlying scenarios, see, e.g., refs.~\cite{Raccanelli_2012, Giannantonio_2008, Raccanelli_2016_cross, Scelfo_2018, Alonso_2020, Ca_as_Herrera_2020, Yang_2023, alonso2024tomographicconstraintsproductionrate}. 
In this work, we focus on a simulation-based approach, and introduce additional complexity into the analysis due to the stochastic nature of different realizations.

%%%%%%%%%%%%%%%%%%%%%%%%%%%%%%%%%%%%%%%%%%%%%%%%%%%%%%%%%%%%%%%%%%%%%%%%%%%%%%%%%%%%%%%%%%%%%%%

\subsection{Intrinsic realization variances}

The statistics of each individual stochastic realization never fully match the underlying theoretical model.
Indeed, the measured angular power spectra~$\hat{C}_\ell$ will be ``scattered'' around the theoretical value~$\tilde{C}_\ell$.
First, we aim to quantify such intrinsic (I) ``scatter'' for realizations of a given theory.
A potential avenue involves defining for each realization~$r$ a~$\chi^2_{I,r}$ statistics given by
\begin{equation}
    \label{eq:snr} 
    \chi_{I,r}^2 \equiv \! \sum_{\ell=2}^{\ell_\mathrm{max}}  \frac{2\ell+1}{2} \mathrm{Tr} \!\left[ \Delta \mathcal{C}_\ell\ \! \qty(\mathcal{C}^{(\rm th)}_\ell)^{-1}\ \!\!\Delta \mathcal{C}_\ell\ \!\! \qty(\mathcal{C}^{(\rm th)}_\ell)^{-1}\ \!\!\right] ,
\end{equation}
where~$\mathcal{C}^{(\rm th)}_\ell$ is a matrix that contains the theoretical angular power spectra, $\mathcal{C}^{(\rm r)}_\ell$ contains its measured value in a given realization, and~$\Delta \mathcal{C}_\ell = \mathcal{C}^{(\rm r)}_\ell-\mathcal{C}^{(\rm th)}_\ell$.
For each angular scale $\ell$, given two galaxy redshift bins~$z_i,z_j$ and GWB frequency bins~$f_p,f_q$, the~$\mathcal{C}_\ell$ matrix is arranged as~\cite{Scelfo_2018, beware1}:

\vspace{4mm}
\makebox[0.78\columnwidth][c]{
  \resizebox{0.5\columnwidth}{!}{ 
\begin{tikzpicture}
% Define the colors
\definecolor{color1}{RGB}{173, 216, 230}
\definecolor{color2}{RGB}{255, 128, 128}
\definecolor{color3}{RGB}{144, 238, 144}

% Size of the squares and separation
\def\square{1.5cm}
\def\sep{0.15cm}

% Fill the squares with colors and add borders

% First row
\fill[color1] (0,0) rectangle (\square,-\square);
\draw[color1, thick] (0,0) rectangle (\square,-\square);

\fill[color1] (\square+\sep,0) rectangle (2*\square+\sep,-\square);
\draw[color1, thick] (\square+\sep,0) rectangle (2*\square+\sep,-\square);

\fill[color2] (2*\square+2*\sep,0) rectangle (3*\square+2*\sep,-\square);
\draw[color2, thick] (2*\square+2*\sep,0) rectangle (3*\square+2*\sep,-\square);

\fill[color2] (3*\square+3*\sep,0) rectangle (4*\square+3*\sep,-\square);
\draw[color2, thick] (3*\square+3*\sep,0) rectangle (4*\square+3*\sep,-\square);

% Second row
\fill[color1] (0,-\square-\sep) rectangle (\square,-2*\square-\sep);
\draw[color1, thick] (0,-\square-\sep) rectangle (\square,-2*\square-\sep);
%\fill[white] (0,-\square-\sep) rectangle (\square,-2*\square-\sep);
%\draw[black, thick] (0,-\square-\sep) rectangle (\square,-2*\square-\sep);

\fill[color1] (\square+\sep,-\square-\sep) rectangle (2*\square+\sep,-2*\square-\sep);
\draw[color1, thick] (\square+\sep,-\square-\sep) rectangle (2*\square+\sep,-2*\square-\sep);

\fill[color2] (2*\square+2*\sep,-\square-\sep) rectangle (3*\square+2*\sep,-2*\square-\sep);
\draw[color2, thick] (2*\square+2*\sep,-\square-\sep) rectangle (3*\square+2*\sep,-2*\square-\sep);

\fill[color2] (3*\square+3*\sep,-\square-\sep) rectangle (4*\square+3*\sep,-2*\square-\sep);
\draw[color2, thick] (3*\square+3*\sep,-\square-\sep) rectangle (4*\square+3*\sep,-2*\square-\sep);

% Third row
\fill[color2] (0,-2*\square-2*\sep) rectangle (\square,-3*\square-2*\sep);
\draw[color2, thick] (0,-2*\square-2*\sep) rectangle (\square,-3*\square-2*\sep);

\fill[color2] (\square+\sep,-2*\square-2*\sep) rectangle (2*\square+\sep,-3*\square-2*\sep);
\draw[color2, thick] (\square+\sep,-2*\square-2*\sep) rectangle (2*\square+\sep,-3*\square-2*\sep);

\fill[color3] (2*\square+2*\sep,-2*\square-2*\sep) rectangle (3*\square+2*\sep,-3*\square-2*\sep);
\draw[color3, thick] (2*\square+2*\sep,-2*\square-2*\sep) rectangle (3*\square+2*\sep,-3*\square-2*\sep);

\fill[color3] (3*\square+3*\sep,-2*\square-2*\sep) rectangle (4*\square+3*\sep,-3*\square-2*\sep);
\draw[color3, thick] (3*\square+3*\sep,-2*\square-2*\sep) rectangle (4*\square+3*\sep,-3*\square-2*\sep);

% Fourth row
\fill[color2] (0,-3*\square-3*\sep) rectangle (\square,-4*\square-3*\sep);
\draw[color2, thick] (0,-3*\square-3*\sep) rectangle (\square,-4*\square-3*\sep);

\fill[color2] (\square+\sep,-3*\square-3*\sep) rectangle (2*\square+\sep,-4*\square-3*\sep);
\draw[color2, thick] (\square+\sep,-3*\square-3*\sep) rectangle (2*\square+\sep,-4*\square-3*\sep);

\fill[color3] (2*\square+2*\sep,-3*\square-3*\sep) rectangle (3*\square+2*\sep,-4*\square-3*\sep);
\draw[color3, thick] (2*\square+2*\sep,-3*\square-3*\sep) rectangle (3*\square+2*\sep,-4*\square-3*\sep);

\fill[color3] (3*\square+3*\sep,-3*\square-3*\sep) rectangle (4*\square+3*\sep,-4*\square-3*\sep);
\draw[color3, thick] (3*\square+3*\sep,-3*\square-3*\sep) rectangle (4*\square+3*\sep,-4*\square-3*\sep);

% Add the text
\node at (0.5*\square, -0.5*\square) {\small $C^{\rm gal,gal}_{\ell,z_i,z_i}$};
\node at (1.5*\square+\sep, -0.5*\square) {\small $C^{\rm gal,gal}_{\ell,z_i,z_j}$};
\node at (0.5*\square, -1.5*\square-\sep) {\small $C^{\rm gal,gal}_{\ell,z_i,z_j}$};
\node at (1.5*\square+\sep, -1.5*\square-\sep) {\small $C^{\rm gal,gal}_{\ell,z_j,z_j}$};

\node at (2.5*\square+2*\sep, -0.5*\square) {\small $C^{\rm gal,GW}_{\ell,z_i,f_p}$};
\node at (3.5*\square+3*\sep, -0.5*\square) {\small $C^{\rm gal,GW}_{\ell,z_i,f_q}$};
\node at (2.5*\square+2*\sep, -1.5*\square-\sep) {\small $C^{\rm gal,GW}_{\ell,z_j,f_p}$};
\node at (3.5*\square+3*\sep, -1.5*\square-\sep) {\small $C^{\rm gal,GW}_{\ell,z_j,f_q}$};

\node at (0.5*\square, -2.5*\square-2*\sep) {\small $C^{\rm gal,GW}_{\ell,z_i,f_p}$};
\node at (1.5*\square+\sep, -2.5*\square-2*\sep) {\small $C^{\rm gal,GW}_{\ell,z_j,f_p}$};
\node at (0.5*\square, -3.5*\square-3*\sep) {\small $C^{\rm gal,GW}_{\ell,z_i,f_q}$};
\node at (1.5*\square+\sep, -3.5*\square-3*\sep) {\small $C^{\rm gal,GW}_{\ell,z_j,f_q}$};

\node at (2.5*\square+2*\sep, -2.5*\square-2*\sep) {\small $C^{\rm GW,GW}_{\ell,f_p, f_p}$};
\node at (3.5*\square+3*\sep, -2.5*\square-2*\sep) {\small $C^{\rm GW,GW}_{\ell,f_p,f_q}$};
\node at (2.5*\square+2*\sep, -3.5*\square-3*\sep) {\small $C^{\rm GW,GW}_{\ell,f_p, f_q}$};
\node at (3.5*\square+3*\sep, -3.5*\square-3*\sep) {\small $C^{\rm GW,GW}_{\ell,f_q, f_q}$};

\node at (-0.5*\square, -2.05*\square-1*\sep) { $\mathcal{C}_\ell = $};
\end{tikzpicture}
}
}
\vspace{4mm}

\noindent This form of the covariance allows to clearly visualize the relation between the observables, and can be extended to any number of frequency and redshift bins. 
The block-diagonal part includes correlations within the same tracer: the top-left block encodes the gal$\times$gal terms between all redshift bins (light blue), while the lower-right one encodes the GWB$\times$GWB terms between all frequency bins  (green).
The off-diagonal blocks encode the cross correlation gal$\times$GWB across different frequency and redshift bins.  

In our simulation-based approach, where we generate galaxy and SMBHBs populations, theoretical angular power spectra and covariance matrices are not known a priori, but have to be deduced from the suite of~$N_r$ realizations.
We estimate the expectation value of the theoretical angular power spectra as
\begin{equation}
    \mathcal{C}^{(\rm th)}_\ell = \bar{\mathcal{C}}_\ell = \frac{1}{N_r} \sum_{j=1}^{N_r} \hat{\mathcal{C}}_{\ell,j} ,
\end{equation}
where the overbar denotes the average value across realizations.
The average value and estimated variance of the~$\chi^2_{I,r}$ for our suite of realizations are defined as
\begin{equation}
    \bar{\chi}_I^2 = \frac{1}{N_r}\sum_r \chi^2_{I,r}, \quad \sigma^2_{\chi^2_I} = \frac{1}{N_r} \sum_r \left( \chi^2_{I,r} - \bar{\chi}^2 \right)^2 \,,
\end{equation}
respectively.
If the angular power spectra are independent Gaussian random variables, the expected value for~$\bar{\chi}^2_I$ and~$\sigma^2_{\chi^2_I}$ can be derived theoretically and is approximately proportional to the number of degrees of freedom of the system, i.e., to~$\ell_\mathrm{max}$.
However, when random variables are correlated, as in the case at hand where~$C^\mathrm{gal,GW}_\ell \neq 0$, the expected value of the average chi-squared and its variance must also be deduced by producing a suite of realizations.

%%%%%%%%%%%%%%%%%%%%%%%%%%%%%%%%%%%%%%%%%%%%%%%%%%%%%%%%%%%%%%%%%%%%%%%%%%%%%%%%%%%%%%%%%%%%%%%

\subsection{Comparing models}
After quantifying the intrinsic variance of our realizations, we can estimate whether a realization from model A is statistically distinguishable from the average realization of a fiducial model B.
For each realization of model A,we define a model comparison (MC)~$\chi^2_{\mathrm{MC},r}$ statistics, defined as
\begin{equation}
    \label{eq:snrAB} 
    \chi^2_{\mathrm{MC},r} = \sum_{\ell=2}^{\ell_\mathrm{max}}  \frac{2\ell+1}{2}  
    \mathrm{Tr}\left[ \Delta \mathcal{C}_\ell\ \qty(\mathcal{C}^{(\rm th, B)}_\ell)^{-1}\ \Delta \mathcal{C}_\ell\  \qty(\mathcal{C}^{(\rm th, B)}_\ell)^{-1}\ \right],
\end{equation}
where~$\Delta \mathcal{C}_\ell = \mathcal{C}^{(\rm r, A)}_\ell - \mathcal{C}^{(\rm th, B)}_\ell$, and each quantity is calculated as described in the previous section.
The value of this new chi-squared statistics is then compared to that of the previous section to assess, with a certain confidence level, whether the deviations from a theoretical model are not compatible with the effect of having different stochastic realizations of the same model.
We then introduce a significance statistics~$D_r$ for model distinguishability for each realization~$r$ as
\begin{equation}
    \label{eq:Ddef}
    {\rm D}_r = \frac{|\chi^2_{\mathrm{MC},r}-\bar{\chi}^2_I|}{\sigma_{\chi^2_I}}.
\end{equation}
Here we consider a LSS-informed GWB as fiducial model, i.e., model B, and assess whether a uniform distribution of SMBHBs can be confidently ruled out as a possible interpretation of the observed signal.
The statistics defined in equation~\eqref{eq:Ddef} allows us to establish that if realization has~$D_r>1, 2, 3$ the uniform model can be excluded at the~$1,2,3\ \sigma$ level.

%%%%%%%%%%%%%%%%%%%%%%%%%%%%%%%%%%%%%%%%%%%%%%%%%%%%%%%%%%%%%%%%%%%%%%%%%%%%%%%%%%%%%%%%%%%%%%%

\section{Results}
\label{sec:results}

%%%%%%%%%%%%%%%%%%%%%%%%%%%%%%%%%%%%%%%%%%%%%%%%%%%%%%%%%%%%%%%%%%%%%%%%%%%%%%%%%%%%%%%%%%%%%%%

\subsection{Loud sources and angular power spectrum distribution}
\label{subsec:gaussianity}

In Figure~\ref{fig:cross_maps} we illustrate that the SMBHB distribution follows the LSS; thus, GWB anisotropies can be LSS-induced.
However, in the context of the GWB, every binary is effectively ``weighted'' by their GW strain, which in turn depends on the astrophysical parameters of the binaries.
Therefore, the statistics of GWB sky maps might be affected by this extra level of stochasticity on top of the stochasticity of spatial distribution of sources we aim to constrain.

\begin{figure}[ht]
  \centerline{
  \begin{subfigure}{0.31\textwidth} 
      \includegraphics[width=\columnwidth]{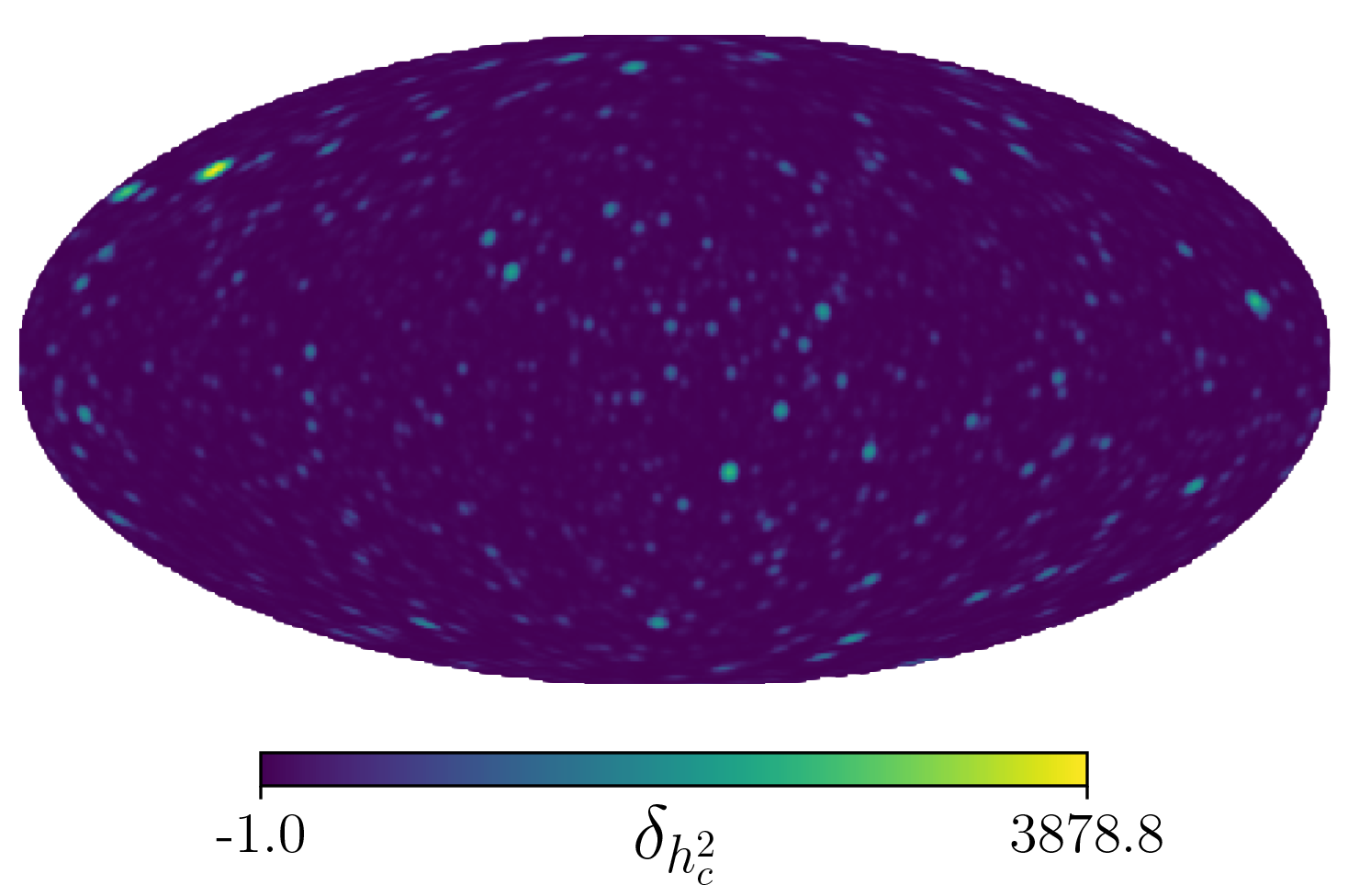}
      \caption{Full GWB, no $h_{\rm res}$ } 
  \end{subfigure}
  \begin{subfigure}{0.31\textwidth} 
      \includegraphics[width=\columnwidth]{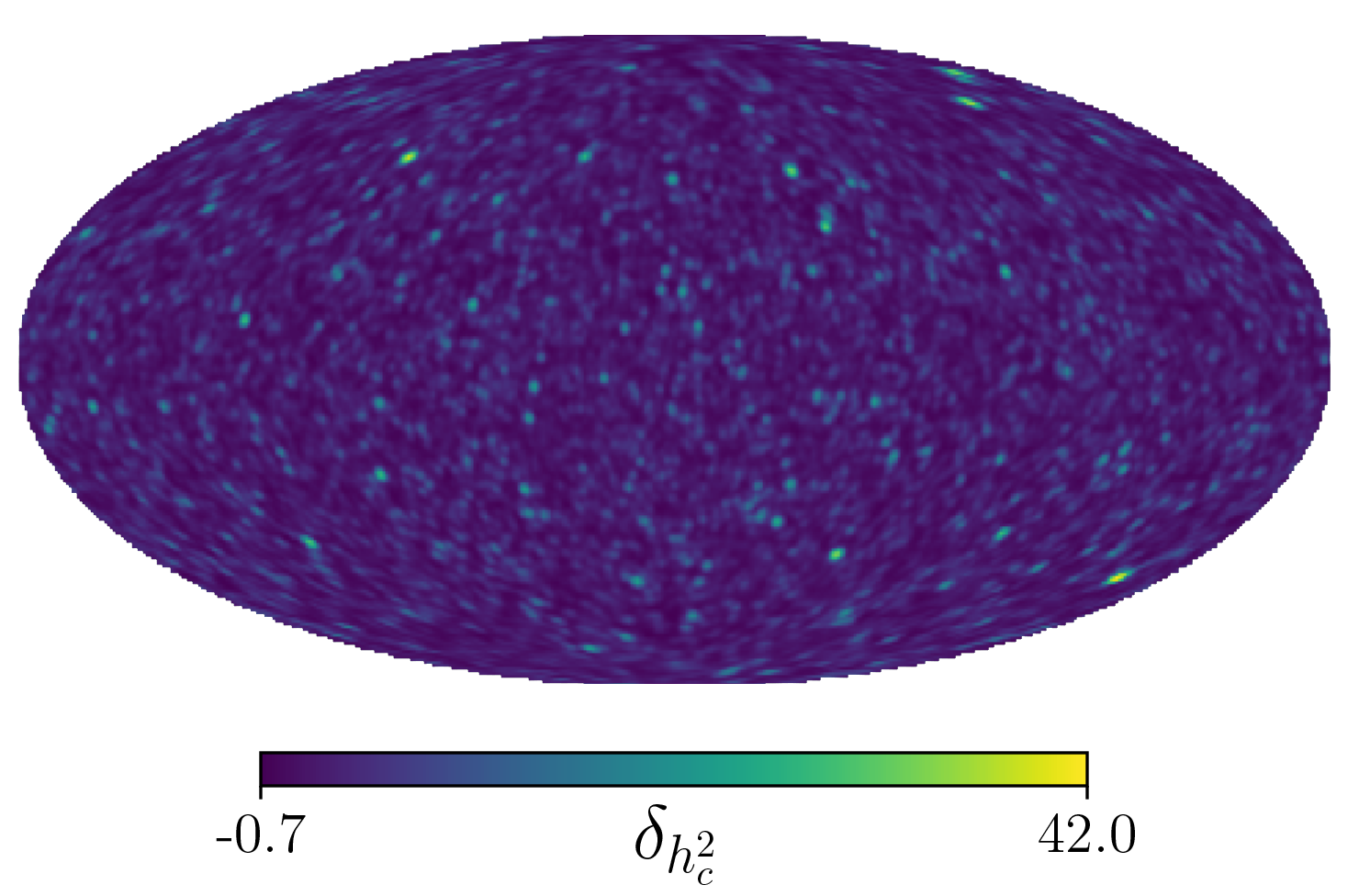}
      \caption{GWB with $h_{\rm res}=10^{-16}$} 
  \end{subfigure}
  \begin{subfigure}{0.31\textwidth} 
      \includegraphics[width=\columnwidth]{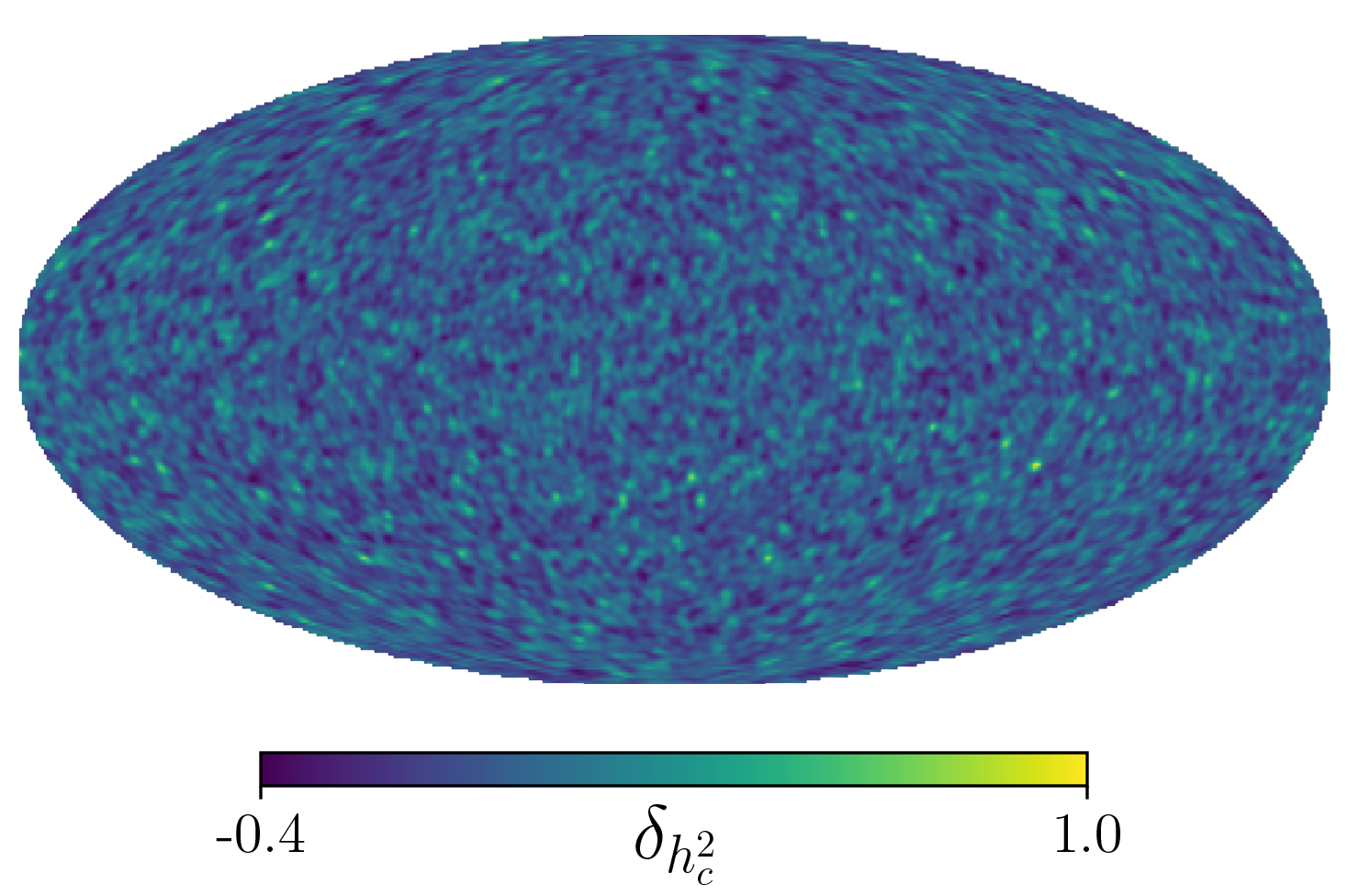}
      \caption{GWB with $h_{\rm res}=10^{-17}$} 
  \end{subfigure}}
  \captionsetup{width=\textwidth}
  \caption{Maps of the GWB characteristic strain overdensities, $\delta_{h_c^2}$. 
  Large-strain sources appear as bright spots, making the $C_\ell$ indistinguishable from a uniform distribution of sources. 
  Moving left to right, we progressively remove the ``loudest'' large-strain sources up to a cut $h_{\rm res}$. 
  By resolving them, anisotropies of the GWB generated by the remaining unresolved population are a representative sample of LSS anisotropies.
  While we use $h_{\rm res}=10^{-16}$ throughout this paper, here we show a $h_{\rm res} =10^{-17}$ map to highlight the potential of future PTA experiments to probe the LSS.}
  \label{fig:maps_brights}
\end{figure}

\begin{figure}[ht]
  \centerline{
  \includegraphics[width=0.65\columnwidth]{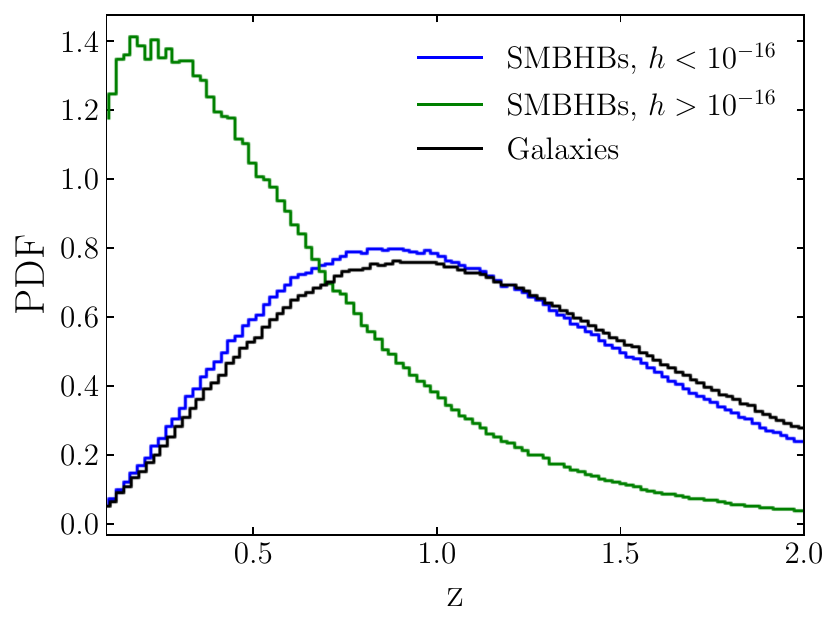}}
  \captionsetup{width=\columnwidth}
  \caption{Loud SMBHBs trace a different redshift distribution than the bulk of SMBHBs which source the GWB. 
  Here we show galaxy and SMBHB redshift distributions, averaged over 1,000 realizations.
  Loud sources are concentrated at low redshifts, where they act as Poisson-like noise, resulting in anisotropies dominated by a small number of sources if not individually resolved and removed. 
  The majority of unresolved SMBHBs creating the GWB spans a larger redshift that traces the underlying galaxy population. 
  Since SMBHBs with $h<10^{-16}$ (blue curve) are a subset of galaxies in our simulations, their distributions are very similar to Galaxies (black curve) but not identical.
  } 
  \label{fig:zhist}
\end{figure}

Figure~\ref{fig:maps_brights} shows the anisotropic GWB sky map obtained from a single realization that includes all sources with emitted strain below a threshold~$h_\mathrm{res}$. 
The threshold~$h_{\rm res}$ is defined as the minimum strain required for a source to be resolved.
These maps represent the idealized spatial distribution of the GWB characteristic strain of the remaining unresolved component.
Although direct pixel-by-pixel reconstruction of such maps by PTAs is challenging, their statistical properties, particularly the angular power spectrum~$C_\ell^{\text{GWB}}$, are accessible through PTA data analysis methods aimed at GWB anisotropy. 
In particular, we observe that for large values of the~$h_\mathrm{res}$ threshold the anisotropies of the maps are totally dominated by the presence of ``loud'', i.e., large strain, sources which appear as bright spots.
These loud sources are also located in galaxies, most likely at low redshifts as showed in figure~\ref{fig:zhist}, thus they still trace the LSS.
However, since anisotropies are completely dominated by a very low number of binaries, their statistic is fundamentally indistinguishable from a uniform, Poisson-like distribution of SMBHBs point sources~\cite{Mingarelli_2013, Ali_Ha_moud_2021, konstandin2024impactcosmicvarianceptas}. 
By resolving and characterizing individual SMBHBs, one can effectively remove their contribution from the unresolved population.
For our SMBHB fiducial model, threshold values of~$h_{\rm res}=10^{-14},\,10^{-15},\,10^{-16}$, correspond to an average number of resolved sources of around $\lesssim 1,\, 30,\, 3600$, respectively.

On the other hand, we observe that by lowering the sensitivity threshold and removing loud sources, we converge towards an actual background scenario, where the global GW emission is the incoherent superposition of many individual unresolved sources, see, e.g., right panel of figure~\ref{fig:maps_brights}, and the statistics becomes more representative of the spatial distribution of galaxies hosting SMBHBs.
Additionally, low strain sources trace a volume of the Universe similar to that probed by the bulk of galaxies, as also showed in figure~\ref{fig:zhist}; therefore, we expect them to be more correlated with LSS.

\begin{figure}[ht]
  \centering
  \includegraphics[width=\columnwidth]{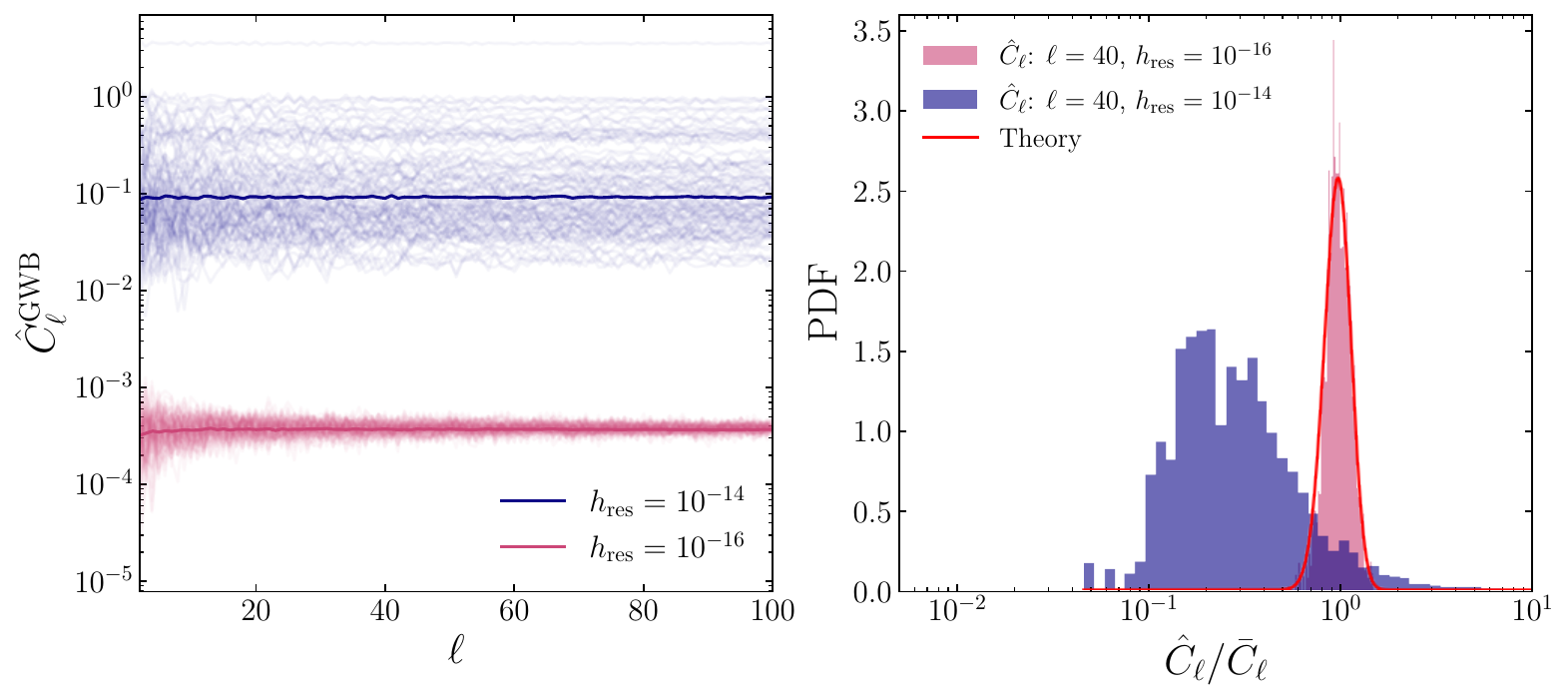}
  \caption{\textit{Left panel}: angular power spectra $\hat{C}_{\ell}$ of 1,000 GWB realizations for two detection thresholds, $h_{\rm res}$. 
  The median~$C_\ell$ value is the solid line. 
  Even when the loudest GW sources are not resolved, they introduce a large variance in the $\hat{C}_{\ell}$ and the statistics are completely Poisson-noise dominated. 
  As we lower the~$h_{\rm res}$ threshold, the variance decreases and the scatter in the~$\hat{C}_{\ell}$ distribution becomes more consistent with cosmic variance.
  \textit{Right panel}: distribution of the~$\hat{C}_\ell$ for a fixed representative angular scale, $\ell=40$, compared to the expected Wishart distribution (both rescaled by their average value for visualization purposes). 
  In the range of angular scales considered in our analysis, the scatter is consistent with the cosmic variance for~$h_{\rm res}\lesssim 8.6\times10^{-16}$, while for higher thresholds the distribution is not compatible with a Wishart distribution within $3\sigma$.}
  \label{fig:cl1416}
\end{figure}

In terms of the statistical properties of GWB sky maps, we observe that when large strain sources are included, maps are effectively shot-noise dominated, and the stochasticity connected to astrophysics induces a large variance between different realizations.
Conversely, removing loud sources we tend towards a situation where the bulk of GWB is effectively given by incoherent superposition of many events, the central limit theorem applies, and different realizations present a degree of anisotropy driven mostly by clustering properties.
These two scenarios are presented in the top panel of figure~\ref{fig:cl1416}, where we show the~$\hat{C}_{\ell}$ of a set of GWB realizations for two thresholds, $h_{\rm res}$.

Regarding angular power spectra, it is well known~\cite{Hamimeche_2008, beware1} that~$\hat{C}_{\ell}$ measured from different Gaussian realizations, i.e., sets of $a_{\ell m}$ harmonic coefficients, of the same underlying theory are statistically distributed according to a Wishart probability distribution function (PDF) $W(2\ell+1,\bar{C}_\ell)$, with $(2\ell+1)$ degrees of freedom and scale given by the mean $\bar{C}_\ell$. 
However, that is not the case when loud sources are included.
We show in the left panel of figure~\ref{fig:cl1416} the comparison between the estimated $C_\ell$s distributions from our suite of realizations, applying different sensitivity cuts.

Additionally, in the right panel of figure~\ref{fig:cl1416}, one can clearly observe that a high threshold for detectability~$h_{\rm res}$ implies a significant deviation with respect to the Wishart distribution for a representative value of the multipole.
In particular, the variance is driven by the high realization dependence of the loudest sources, both in number and spatial position.
We perform a Kolmogorov–Smirnov test~\cite{smirnov} to establish at which threshold the field has a two-point statistics that is compatible with a Wishart. 
We find that for~$h_{\rm res}\gtrsim 8.6\times10^{-16}$, the~$\hat{C}_\ell$ distribution is not consistent with a Wishart PDF within~$3\sigma$ in the range of~$\ell$ modes we consider in the following sections, see also appendix~\ref{app:ks_test} for additional details. 
This implies that the two-point function alone is not sufficient to describe the statistics of the field.
This is especially relevant at the current sensitivity level of PTA experiments, which currently do not resolve any single GW sources.
Therefore, in the following sections, we will perform the statistical analysis only for sensitivity cuts below this GWB strain limit, where GWB maps are dominated by a continuous distribution of sources, for which our Gaussian formalism provides a good description of the statistics of the field.

%%%%%%%%%%%%%%%%%%%%%%%%%%%%%%%%%%%%%%%%%%%%%%%%%%%%%%%%%%%%%%%%%%%%%%%%%%%%%%%%%%%%%%%%%%%%%%%%%%%%%%%%%

\subsection{GWB$\times$LSS Cross-correlation}
\label{subsec:crss-corr}

For the purpose of maximizing the LSS-induced anisotropies, we create a single redshift and frequency bin for galaxies and GWB, respectively, and we consider the case where~$h_{\rm res}=10^{-16}$.
We find comparable results when including multiple redshift bins, and discuss the prospects of frequency tomography in section~\ref{sec:discussion}.
For each realization, we obtain the GWB two-point statistics produced by SMBHBs uniformly distributed on the sphere by creating sub-realizations where SMBHBs angular positions are drawn from a uniform pdf, and by averaging over this new set of~$C_\ell$s. 
The obtained angular power spectrum is consistent with a flat shot-noise spectrum.
Finally, the analysis of section~\ref{sec:analysis} is performed both by excluding and including the cross-correlation signal, i.e., by setting the signal to zero in the corresponding entry of the covariance matrix, to further emphasize its constraining power.

\begin{figure}[ht]
  \centering
  \includegraphics[width=\textwidth]{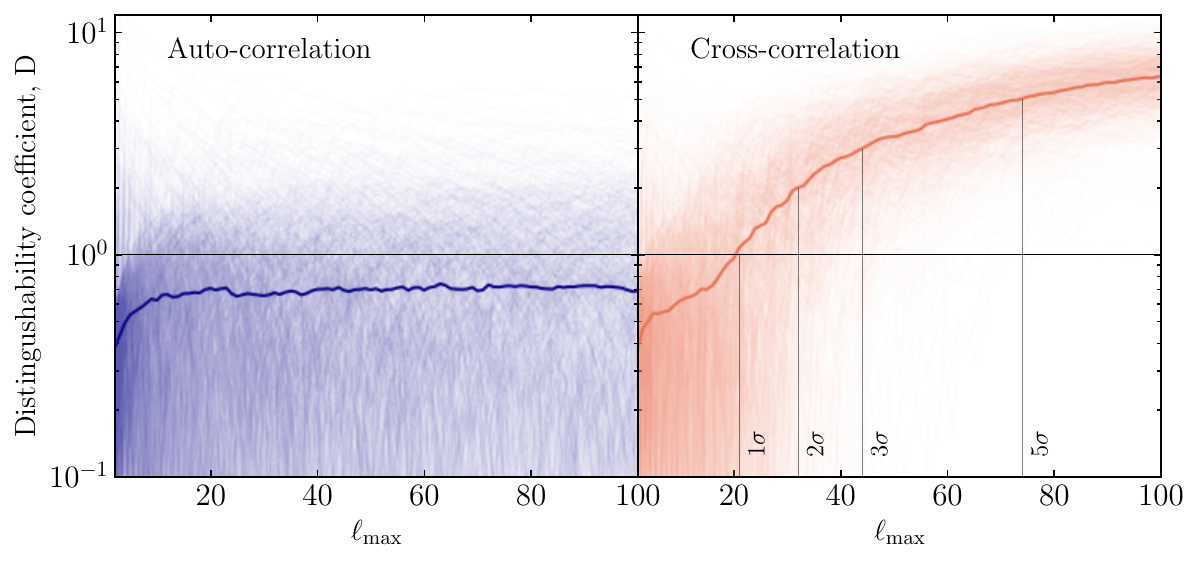}
  \caption{LSS becomes measurable via cross-correlation searches in PTA experiments at a $1,2,3,5\sigma$ level ($ {\rm D}=1,2,3,5$) for $\ell_{\rm max}\geq 20, 30, 42, 72$, respectively, where ${\rm D}$ is the distinguishability coefficient. 
  Each realization is plotted in light shade with a dark median. 
  We assume a resolvability threshold of $h_{\rm res}\geq 10^{-16}$ and a maximum frequency of $f_{\rm max}=6 \,{\rm nHz}$. 
  \textit{Left}: Using just the auto-correlation, LSS-induced anisotropies are indistinguishable from the ones sourced by a uniform distribution of sources.
  \textit{Right}: A cross-correlation analysis provides ${\rm D}\geq1$ for $\ell_{\rm max}\gtrsim 20$.} 
  \label{fig:auto_cross}
\end{figure}

We show in figure~\ref{fig:auto_cross} the D factor for auto-correlation only and cross-correlation analysis as a function of the maximum resolution of the PTA experiment.
In each panel, we show the statistics of each realization (either directly or by drawing a band that encloses them) and the median of the distribution.
Despite having removed loud sources, using only the information contained in the GWB auto-correlation does not allow us to discriminate between different SMBHBs spatial distributions, as showed in the top-left panel of the Figure.
However, when cross-correlations are included in the analysis, we can distinguish between the two scenarios at more than one sigma confidence when~$\ell_{\rm max}\gtrsim 20$ and $h_{\rm res}=10^{-16}$, i.e., in a SKA-like survey.

Loud sources are expected to be scattered throughout the entire frequency range, see figure~\ref{fig:smbhb_hists}, and their contribution to the characteristic strain is greater for high frequency sources, as we can deduce from equation~\ref{eq:hc2}.
In a single-bin scenario, we then focus on a narrower frequency range to remove the high-frequency tail of the distribution that creates large overdensities in the GWB maps.
Specifically, we stack information from the first two NANOGrav frequency bins up to~$6$ nHz.
We also compute our results for varying~$f_{\max}$, and find that the D coefficient at~${\ell_{\rm max}=70}$ decreases as~${{\rm D}_{10 {\rm nHz}}\simeq 0.86\, {\rm D}_{6 {\rm nHz}}}$ and~${{\rm D}_{20 {\rm nHz}}\simeq 0.6\, {\rm D}_{6 {\rm nHz}}}$, respectively.

Importantly, we find that the ability to distinguish between different SMBHB spatial distributions improves significantly with increasing~$\ell_{\max}$.
We achieve~$1$, $2$, $3$, and~$5\sigma$ significance for~$\ell_{\max} \gtrsim 20, 30, 42, 72$, respectively. 
This demonstrates that high angular resolution in PTA experiments is critical for extracting anisotropies imprinted by LSS.
The combined advances provided by SKAO, in terms of both number of pulsars and isotropic response, will make detection of GWB anisotropies at cosmologically relevant scales feasible within SKA2's operational timeline.

The significance of this result depends on how strongly SMBHBs are clustered.
Standard expectations suggest that larger biases necessarily correspond to a stronger distinguishability from the isotropic case.
Given the theoretical uncertainty associated with the precise value of the GWB bias, in figure~\ref{fig:bias_dependence} of  appendix~\ref{app:gwb_bias_detectability} we estimate the value of the D metric for two additional cases corresponding to the scenarios where the value of the bias is twice and half the fiducial value obtained from simulations, confirming this expectation.

Finally, a key factor in reaching these angular resolutions will be the effective identification and removal of loud sources, which we have demonstrated is essential for accessing the underlying LSS distribution in the GWB.
We note that there is an intrinsic trade-off between the source-removal threshold $h_{\rm res}$ and the maximum usable $\ell_{\rm max}$: more stringent source-removal thresholds (lower $h_{\rm res}$) require sensitivity to smaller angular scales to successfully identify and subtract the corresponding sources, coupling the two parameters in practice.
Nevertheless, our results imply that a realization of the GWB can provide a clear imprint of LSS anisotropies, and a cross-correlation approach is crucial to extract this signal.
This further highlights the importance of resolving continuous GW sources in the GWB, leaving the unresolved population as a fair sample of the underlying LSS, and a linear overdensity field description to be a suitable statistic for a cosmological anisotropy analysis.

%%%%%%%%%%%%%%%%%%%%%%%%%%%%%%%%%%%%%%%%%%%%%%%%%%%%%%%%%%%%%%%%%%%%%%%%%%%%%%%%%%%%%%%%%%%%%%%%%%%%%%%%%

\section{Summary and Discussion}
\label{sec:discussion}

Cross-correlation analysis between the GWB and LSS is a critical tool to unravel the complex nature of the GWB. 
Both galaxies and SMBHBs trace the same underlying dark matter distribution, but through different astrophysical processes and with different biases. 
Cross-correlation effectively isolates their common source—the LSS—while mitigating tracer-specific systematics and noise properties, as is the case for galaxies and the GWB. 
If two tracers can be considered as independent samples of the same field, their cross-correlation becomes entirely free of shot noise. 
Although a residual correlated shot noise can exist in the cross-spectrum if the tracers are not fully independent~\citep{cusin2025measuringanisotropiesptaband}, its contribution in the case at hand is generally negligible compared to the clustering signal at the scales of interest and is inherently accounted for in our simulation-based covariance estimation.
In the case of the GWB, auto-correlation is particularly limited by the low number of sources.
Cross-correlation with galaxies effectively provides a unique smoking gun of the LSS imprint in the signal, and, in the case where galaxies have known redshift, it can be employed for a clustering-based reconstruction of the redshift distribution of SMBHBs~\citep{ménard2014clusteringbasedredshiftestimationmethod, Kovetz_2017}.

In this work, we show how the GWB generated by SMBHBs is an effective tracer of the LSS of our Universe.
We produced 1,000 realizations of SMBHB catalogs directly from dark matter~$N$-body simulations. 
We then built full-sky overdensity maps of galaxy number counts and GWB characteristic strain, without relying on a parametrized model of the angular power spectra, as, for instance, done by ref.~\cite{sah2024}.
We investigate the impact of loud GW sources on the statistics of the GWB anisotropies, proving that for~$h_{\rm res}\gtrsim 10^{-15}$ the~$C_\ell$ distribution is not compatible with a Wishart distribution within~$3\sigma$. 
This means that~$C_\ell$s are not the best suited summary statistic for the GWB. 
Indeed, this is very relevant for current PTA experiments, since it implies that the central limit theorem does not apply to GWB maps; thus, the~$C_\ell$ do not fully capture the information content and features of the GWB.

We also showed that using our method, given one realization of the GWB, such as the one we inhabit, we can robustly distinguish between a GWB with an imprint of the LSS and uniform distribution of SMBHBs at more than~$3\sigma$ confidence with a next-generation PTA experiment such as SKAO~\cite{Xin_2021}.
This is achievable if sources with strain larger than~$h_{\rm res}\gtrsim 10^{-16}$ are resolved, which is also feasible with the SKAO~\cite{Xin_2021}.
The LSS imprint can be detected at~$1\sigma$, $2\sigma$, $3\sigma$, and~$5\sigma$ significance levels for maximum angular resolutions of $\ell_{\max} = 20, 30, 42$, and~$72$, respectively.
These significance levels represent the potential sensitivity assuming that GWB characteristic strain maps can be reliably reconstructed up to these angular resolutions and that the dominant shot noise from the brightest individual SMBHBs is effectively removed. 
Actual sensitivities achievable by real PTA experiments will depend on the full characterization and mitigation of instrumental noise, pulsar timing irregularities, and map-making uncertainties, aspects which are subjects of ongoing research and planned for our future work.
Furthermore, our work shows that an approach based only on the GWB auto-correlation does not allow to distinguish between SMBHB distribution models, while a cross-correlation analysis does.
Recently,~\cite{grimm2024, allen2024sourceanisotropiespulsartiming} also found that the Poisson noise contribution from a finite number of SMBHBs has a much larger impact on deviations from Hellings and Downs correlations~\cite{HD83}  -- used to detect an isotropic GWB -- than anisotropies from the LSS.

A detection of the cross-correlation signal opens the possibility of measuring the GWB bias since~$C_\ell^{\rm gal,GW} \propto b_{\rm gal}\, b_{\rm GW}\, C_\ell^{mm}$.
The statistical metric~$D$ defined in this work has been used to compare the hypothesis of a GWB with an LSS imprint against the null hypothesis of a uniformly-distributed (isotropic) GWB. 
As such, our framework does not directly constrain bias models or measure~$b_{\rm GW}$; rather, it quantifies whether the cross-correlation signal is distinguishable from zero.
Estimating the error in the potential measurement of~$b_\mathrm{GWB}$ would require a different setup, either performing a Fisher analysis from simulations or a full Bayesian parameter estimation.
In fact, ref.~\cite{Sah:2025uuk} recently showed with a Fisher forecast in an analytical setup how the cross-correlation spectrum is a powerful tool to constrain population and evolution history parameters of the SMBHB population.
However, the computational cost of these procedures mentioned above, which inherently require the exploration of the parameter space of our model, makes the task computationally challenging.
Taking a simplistic approach, we could loosely use the confidence level associated to~$D$ as a first estimate of the error on the GWB, for example by transforming a~$3\sigma$ confidence level distinguishability into~$\sigma_{b_\mathrm{GWB}} \approx b_\mathrm{GWB}/3 \approx 0.8$.
Despite not being a statistically robust procedure, this kind of estimate can be helpful to ``guide our eyes'' while analysing the scientific potential of this probe.

Although ref.~\cite{yang2024anisotropynanohertzgravitationalwave} also explored how SMBHB clustering trace the LSS, they did not carry out a cross-correlation analysis.
Furthermore, refs.~\cite{sah2024, sah2024_imprints} developed a theoretical formalism to forecast cross-correlation signals between the GWB and LSS; however, their method cannot accurately predict the impact of the discreteness of the SMBHB population.
We show that this is particularly relevant for the loudest low-redshift sources, with important implications on the GWB~$C_\ell$ and the cross-correlation signal -- that is to say that the~$C_\ell$s fail to capture the features of the GWB, as we explained above. 

We now consider the accuracy and limitations of our methods. 
Here we consider galaxy clustering with a single redshift bins; however, a tomographic approach might allow us to better capture the underlying properties of the population that sources the GWB.
Such an approach would require the generation of more accurate lightcones that include both redshift-space distortions~\cite{kaiser, Hamilton_1998, Raccanelli_2013} and projection effects~\cite{Bonvin11, Jeong_2012, Challinor_2011, bertacca_beyond_2012, Bertacca_2018} to obtain an unbiased estimates of the cross-bin angular power spectra~$C^{\rm gal,gal}_{\ell,z_i,z_j}$~\cite{Matsubara_1997, Yoo_2009, raccanelli2016doppler, Dio_2016, semenzato2024SFB, spezzati20243d}.
Moreover, by populating the lightcone with different types of galaxies, we will be able to better understand the nature of the cross-frequency power spectra~$C^{\rm GW,GW}_{\ell,f_p,f_q}$. 
This will ultimately help to design an optimal frequency binning strategy in such a cross-correlation search.

Our efforts in this work were focused on predicting what the GWB signal looks like in nature, if LSS imprints would ever be detectable, and how to approach their detection. 
However, we did not simulate what this signal would look like in PTA data. 
In a future study, we plan to do this by adding pulsar noise~\cite{nanograv_noise,larsen2024}, accounting for the anisotropic distribution of pulsars on the sky~\cite{alihaimoud2020, Ali_Ha_moud_2021}, as well as the way the GWB is detected~\cite{HD83, Mingarelli_2013, cornish2014}. All these improvements will help us to better model the expected signal in our cross-correlation searches and improve our abilities to detect it.

Although our analysis incorporates realistic astrophysical populations and detection thresholds informed by PTA sensitivity curves, we have not implemented a full map-reconstruction pipeline from simulated timing residuals. 
Such a pipeline would primarily constrain the maximum angular resolution~($\ell_{\max}$) achievable in the reconstruction, essentially imposing a resolution limit consistent with the capabilities of a given experiment and related to the number of available pulsar pairs. 
The process of reconstructing a GWB map (or its $C_\ell$s) from timing data is imperfect and introduces its own noise and potential bias, which effectively adds to the variance of the measured~$C_\ell$ and can limit the achievable~$\ell_{\rm max}$. 
While some analytical formalisms for these effects assume a continuous limit~\cite{Hotinli_2019} (e.g., infinite pulsars), detailed knowledge of pulsar locations and noise can inform optimal map-making strategies, see, e.g., ref.~\cite{alihaimoud2020}.
Incorporating these instrumental effects would likely reduce the significance of the LSS imprint detection compared to our current idealized forecasts. 

Our results therefore represent an estimate of the \textit{potential physical information content} available, motivating further development of optimized analysis pipelines. 
Our current approach offers significant advantages in isolating the intrinsic physical limitations from observational constraints, though we acknowledge certain trade-offs. 
The SNR values presented here represent theoretical upper bounds that would likely be moderated in a full Bayesian analysis accounting for additional systematics, reconstruction errors, and confusion noise. 
The removal of loud individual continuous wave sources remains a critical consideration in GWB anisotropy detection, as their spectral leakage can contaminate multiple angular scales. 
However, precise localization of bright sources, potentially through their electromagnetic counterparts, can substantially improve the recovery of the underlying GWB anisotropy structure.

We also plan to employ our general framework to test the connection between GWB anisotropies and the underlying SMBHB population properties, especially in light of the existence of low-redshift, loud sources located in the local Universe, whose existence appears to be in disagreement with EM observations~\cite{satopolito2023nanogravsbigblackholes, satopolito2024distributiongravitationalwavebackgroundsupermassive}.
Our new pipeline allows the construction of realistic GWBs directly from dark matter simulations, which can be implemented in a full Bayesian data analysis. 
This will in turn provide an avenue to directly constrain the physics driving the SMBHB mergers and, eventually, cosmological parameters themselves.

In summary, this work highlights the importance of cross-correlation techniques in characterizing and analyzing the GWB, effectively making it another tracer of LSS.
In fact, our technique can be used to explore the possible existence of a cosmological GWB component~\cite{Alba_2016, Contaldi_2017, Bartolo_2020, Valbusa_Dall_Armi_2021, Afzal_2023}, in addition to an astrophysical one. 
Indeed, since cosmological GWB anisotropies share common statistical features with, e.g., the Cosmic Microwave Background~\cite{Ricciardone_2021}, it would be interesting to explore, e.g., a~$3\times 2$pt function cross-correlation analysis between the CMB, the LSS, and the GWB.
As observational techniques continue to improve and more data become available, this novel framework will allow us to explore fundamental physics not accessible by any other means.

%%%%%%%%%%%%%%%%%%%%%%%%%%%%%%%%%%%%%%%%%%%%%%%%%%%%%%%%%%%%%%%%%%%%%%%%%%%%%%%%%%%%%%%%%%%%%%%%%%%%%%%%%

\section*{Acknowledgements}
The authors would like to thank Bjorn Larsen, Eleonora Vanzan, Sarah Libanore, Sabino Matarrese, Nikhil Padmanabhan, and Deepali Agarwal for useful discussions and comments on the draft.
This work is partly supported by ICSC - Centro Nazionale di Ricerca in High Performance Computing, Big Data and Quantum Computing, funded by European Union - NextGenerationEU.
CMFM acknowledges support from the National Science Foundation from Grants NSF PHY-2020265 and AST-2414468. 
This work was also supported by the Flatiron Institute, part of the Simons Foundation.
AR acknowledges funding from the Italian Ministry of University and Research (MIUR) through the ``Dipartimenti di eccellenza'' project ``Science of the Universe''.
NBe is supported by PRD/ARPE 2022 ``Cosmology with Gravitational waves and Large Scale Structure - CosmoGraLSS''.
DB acknowledges partial financial support from the COSMOS network (www.cosmosnet.it) through the ASI (Italian Space Agency) Grants 2016-24-H.0, 2016-24-H.1-2018 and 2020-9-HH.

%%%%%%%%%%%%%%%%%%%%%%%%%%%%%%%%%%%%%%%%%%%%%%%%%%%%%%%%%%%%%%%%%%%%%%%%%%%%%%%%%%%%%%%%%%%%%%%%%%%%%%%%%
  
\appendix

\section{Galaxy mock catalog}
\label{app:galaxy_mock}

\subsection{HOD}
The baseline \Abacus{AbacusHOD} parametrization~\cite{HOD_Yuan:2021izi} describes the mean expected number of central and satellite galaxies per halo as a function of halo mass with five parameters~$\{M_\mathrm{cut}, M_{1-\mathrm{sat}}, \sigma_M, \alpha_M, \kappa\}$ as
\begin{equation}
    \begin{aligned}
        N_{\mathrm{cent}}^{\mathrm{LRG}}(M) & = \frac{1}{2}\mathrm{erfc} \left[\frac{\log_{10}(M_{\mathrm{cut}}/M)}{\sqrt{2}\sigma_M}\right]\,, \\
        N_{\mathrm{sat}}^{\mathrm{LRG}}(M) & = \left[\frac{M-\kappa M_{\mathrm{cut}}}{M_1}\right]^{\alpha_M} N_{\mathrm{cent}}^{\mathrm{LRG}}(M)\,.
    \end{aligned}
\end{equation}
Here, $M_{\mathrm{cut}}$ and~$\kappa M_\mathrm{cut}$ are the minimum halo masses to host a central and satellite galaxy, respectively, $M_1$ is the characteristic halo mass that hosts one satellite galaxy, $\alpha_M$ is the power law index, and~$\sigma_M$ characterizes the transition tilt in the number of central galaxies.
The model enforces each halo to host at most one central galaxy, which is placed at the center of mass of the largest sub-halo.
The number of satellite galaxies is drawn from a Poisson distribution with mean~$\bar{n}^{\mathrm{LRG}}_{\mathrm{sat}}(M)$, and they are assigned to particles within the halo with equal weights. 

%%%%%%%%%%%%%%%%%%%%%%%%%%%%%%%%%%%%%%%%%%%%%%%%%%%%%%%%%%%%%%%%%%%%%%%%%%%%%%%%%%%%%%%%%%%%%%%%%%%%%%%%%

\subsection{Stellar masses}

\begin{figure}[ht]
    \centering
    \includegraphics[width=\textwidth]{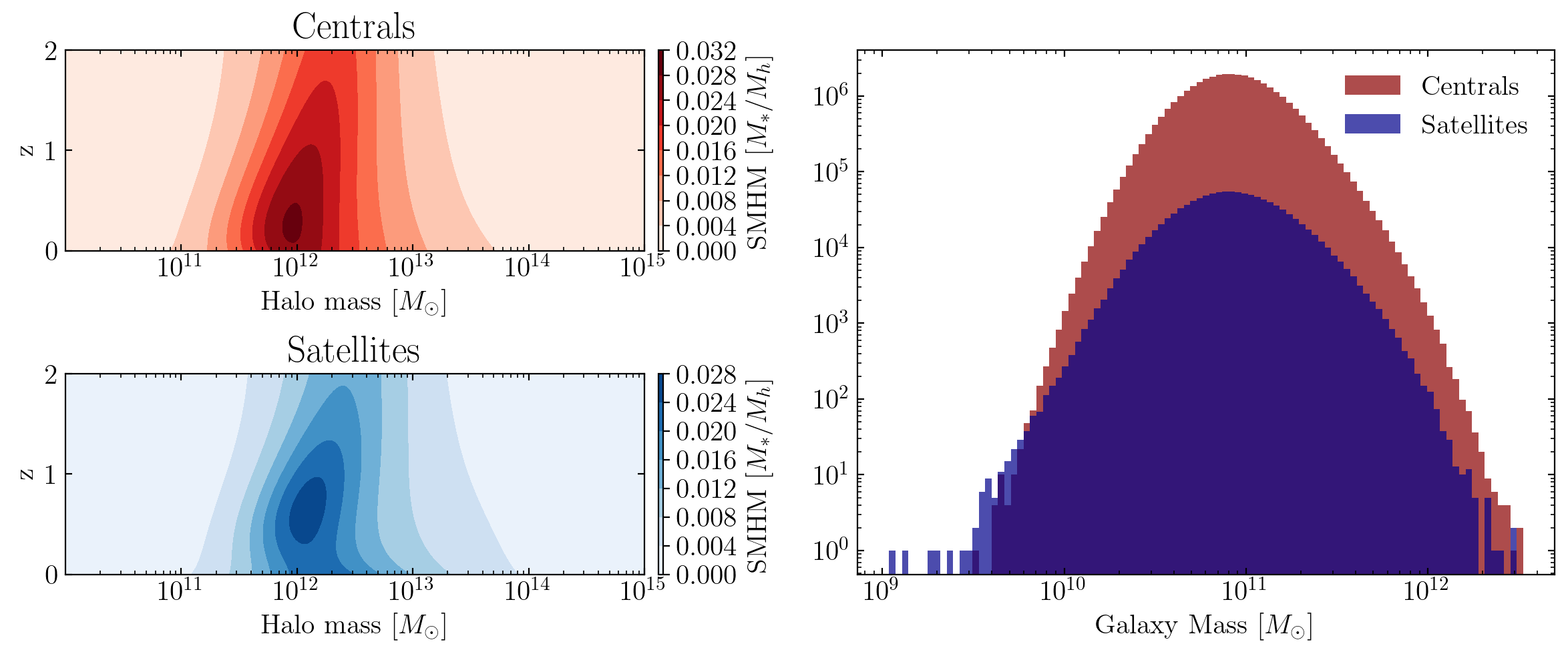}
    \caption{\textit{Left panel:} stellar mass to halo mass relation for central and satellite galaxies for the redshift range considered in this work.
    \textit{Right panel:} galaxy mass histogram distribution for central and satellite galaxies.}
\label{fig:SMHM} 
\end{figure}

We use the stellar mass~$M_*$ to halo mass~$M_h$ ratio (SMHM ratio) of the \textsc{UniverseMachine} approach~\cite{universe_machine_Behroozi:2019kql} to assign stellar masses to galaxies.
The~\textsc{UniverseMachine} framework parametrizes the galaxy star formation rate as a function of the halo potential well depth, redshift, and assembly history, providing a self-consistent fit to experimental data by connecting the star formation history to the properties of the host.
The SMHM fits are provided for both central and satellite galaxies, including redshift dependence (although redshift evolution is not significant for redshifts $z\lesssim 5$). 
The SMHM relation is parametrized as a double power-law plus a Gaussian function of the form 
\begin{equation}
    \log_{10} \left( \frac{M_\ast}{M_1} \right) = \epsilon - \log_{10} \left( 10^{-\alpha x} + 10^{-\beta x} \right) + \gamma \exp \left[ -0.5 \left( \frac{x}{\delta} \right)^2 \right]\,,
\end{equation}
where~$x = \log_{10}(M_h/M_1)$.
The parameters~$\epsilon, \alpha, \beta, \gamma, \delta$ are functions of redshift modeled as a
\begin{equation}
    \begin{aligned}
        \log_{10} \left(\frac{M_1}{\Msun}\right) &=  M_0 + M_a (a-1) - M_\mathrm{loga} \log(a) + M_z z, \\
        \epsilon &= \epsilon_0 + \epsilon_a (a-1) - \epsilon_\mathrm{loga} \log(a) + \epsilon_z z, \\
        \alpha &= \alpha_0 + \alpha_a (a-1) - \alpha_\mathrm{loga} \log(a) + \alpha_z z, \\
        \beta &= \beta_0 + \beta_a (a-1) + \beta_z z, \\
        \delta &= \delta_0, \\
        \log_{10}(\gamma) &= \gamma_0 + \gamma_a (a-1) + \gamma_z z, \\
    \end{aligned}
\end{equation}
where $a = 1/(1+z)$ is the cosmological scale factor, and the values of the fitting parameters are provided in Table J1 of~\cite{universe_machine_Behroozi:2019kql}.
The stellar mass is then assigned to the galaxies by drawing from the SMHM relation, also accounting for the scatter in the relation to introduce some stochasticity in the stellar mass assignment.
Figure~\ref{fig:SMHM} summarizes the mass-assignment distribution for both central and satellite galaxies. The left panel shows the SMHM relation as a function of halo mass and redshift, while the right panel shows the final mass distribution for the two sub-populations. 

%%%%%%%%%%%%%%%%%%%%%%%%%%%%%%%%%%%%%%%%%%%%%%%%%%%%%%%%%%%%%%%%%%%%%%%%%%%%%%%%%%%%%%%%%%%%%%%%%%%%%%%%%

\subsection{Validation of galaxy angular power spectra}
\label{app:class_validation}

\begin{figure}[ht]
    \centering
    \includegraphics[width=0.6\columnwidth]{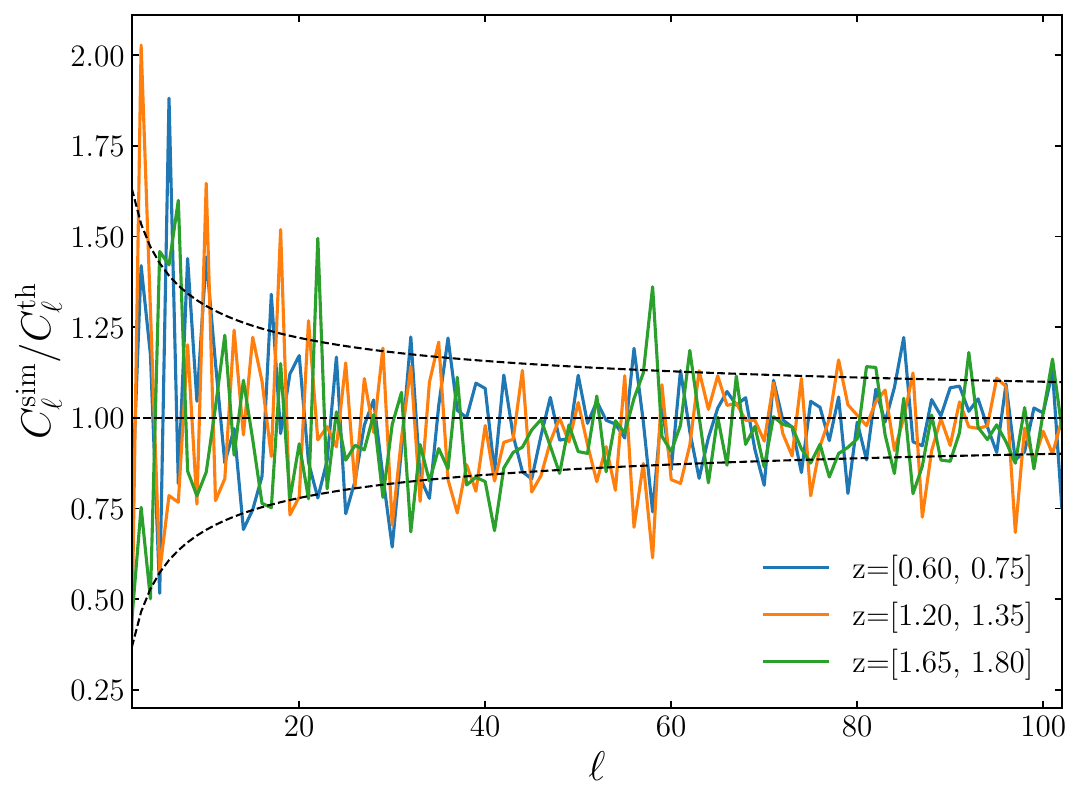}
    \caption{Ratio of the simulated galaxy angular power spectrum to theoretical prediction for three representative redshift bins. 
    Dashed black lines indicate the $1\sigma$ cosmic variance uncertainty band.
    This results show that the simulated galaxy clustering is consistent with linear theory predictions within the expected scatter due to cosmic variance.}
    \label{fig:class_comparison}
\end{figure}

Here we compare the galaxy angular power spectra with the theoretical prediction of linear perturbation theory.
First, we use the galaxy and matter fields from \Abacus{AbacusSummit} to estimate the galaxy bias in thin redshift slices as~$b(z)=\sqrt{C_{\ell}^{\rm gal,gal}(z)/C_{\ell}^{\rm mm}(z)}$.
We find that the sample has a linear bias of~$b_\mathrm{gal}\in [1.6,4.1]$ between redshift~$[0.2,2.0]$, with an average bias of the entire sample of~$b_\mathrm{gal}\simeq 2.4$, which is in agreement with ``standard'' estimates of galaxy population properties.
Then, we obtain the galaxy angular power spectrum from  \textsc{CLASS}~\cite{lesgourgues2011cosmiclinearanisotropysolving} for a population of galaxy with the same redshift distribution and bias of our mock.
Finally, we build the theoretical prediction by adding the shot noise angular power spectrum to CLASS result.
In figure~\ref{fig:class_comparison}, we show the ratio of the simulated galaxy angular power spectra spectrum to the theoretical prediction for three representative redshift bins.
For the scales relevant in our analysis, i.e., $\ell\lesssim 100$, the ratio is consistent with unity within the expected cosmic variance uncertainty, confirming that our galaxy mock is consistent with linear theory predictions given its effective bias and number density.

%%%%%%%%%%%%%%%%%%%%%%%%%%%%%%%%%%%%%%%%%%%%%%%%%%%%%%%%%%%%%%%%%%%%%%%%%%%%%%%%%%%%%%%%%%%%%%%%%%%%%%%%%

\section{Validation of the galaxy-GW and the GW-GW angular power spectra}
\label{app:galaxygw_gwgw_validation}

\begin{figure}[ht]
    \centering
    \includegraphics[width=\columnwidth]{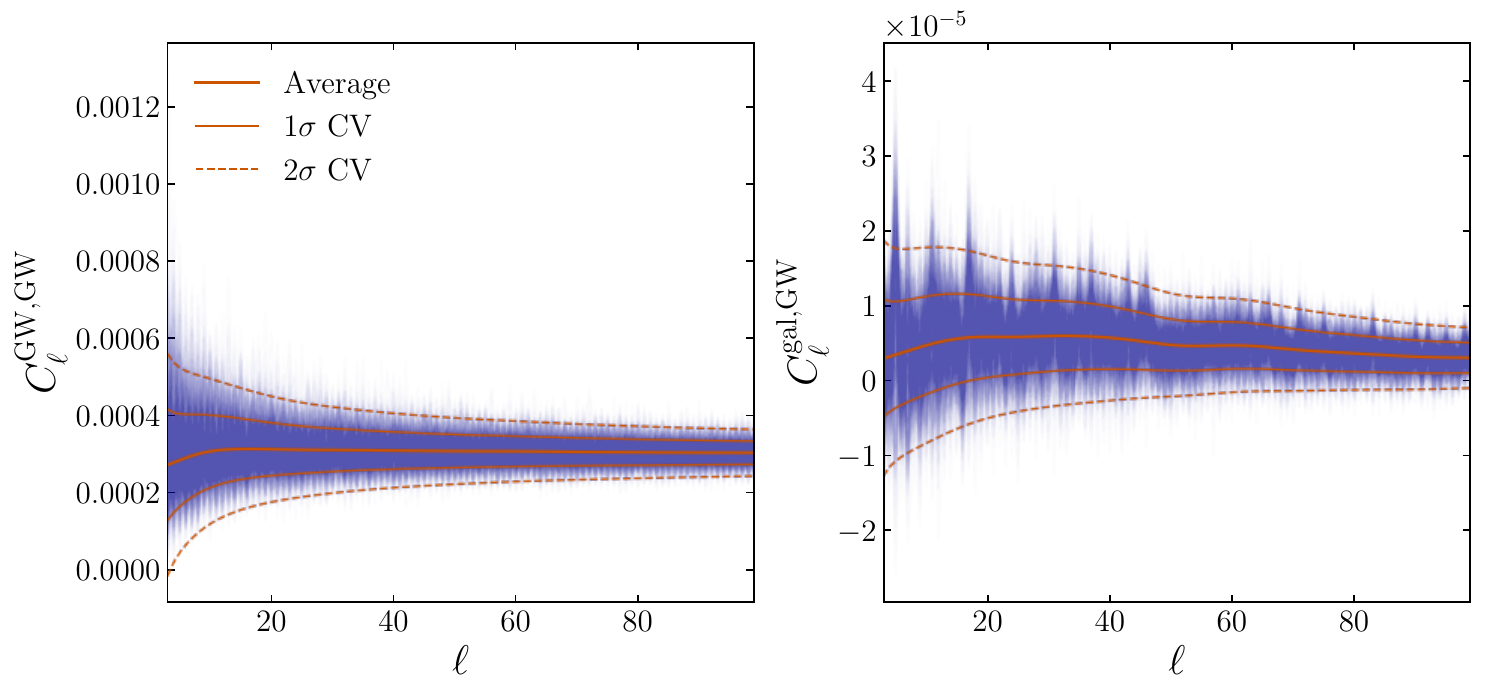}
    \caption{\justifying Set of 1,000 GW-GW (\textit{left panel}) and galaxy-GW (\textit{right panel}) angular power spectra obtained from simulations. 
    Thick, thin, and dashed dark orange lines indicate the average of the realizations, and the $1\sigma$ and~$2\sigma$ cosmic variance uncertainty band.
    This results show that the simulated angular power spectra are scattered in accordance with the expected level due to cosmic variance.}
    \label{fig:ggw_gwgw_validation}
\end{figure}

In this appendix, we report our validation tests of the galaxy-GW and GW-GW angular power spectra.
We show in figure~\ref{fig:ggw_gwgw_validation} the set of galaxy-GW and GW-GW angular power spectra obtained from our set of 1,000 realizations, along with the average and expected cosmic variance error bands.
As we observe from the figure, both categories of power spectra present a statistical distribution compatible with the theoretical cosmic variance scatter.
Similarly to what has been done in appendix~\ref{app:class_validation}, we use the average galaxy-GW angular power spectrum to provide an estimate of the GWB bias as~$b_\mathrm{GWB}\simeq C^\mathrm{gal,GW}_\ell/(b_\mathrm{gal}C^{mm}_\ell) \simeq 2.6$, where in this case we used the average bias~$b_\mathrm{gal}=2.4$.

%%%%%%%%%%%%%%%%%%%%%%%%%%%%%%%%%%%%%%%%%%%%%%%%%%%%%%%%%%%%%%%%%%%%%%%%%%%%%%%%%%%%%%%%%%%%%%%%%%%%%%%%%

\section{Complementary details on Statistics}
\label{app:covariance}

%%%%%%%%%%%%%%%%%%%%%%%%%%%%%%%%%%%%%%%%%%%%%%%%%%%%%%%%%%%%%%%%%%%%%%%%%%%%%%%%%%%%%%%%%%%%%%%%%%%%%%%%%

\subsection{Intrinsic and model-comparison chi-squared from simulations}

\begin{figure}
\centering
\includegraphics[width=\textwidth]{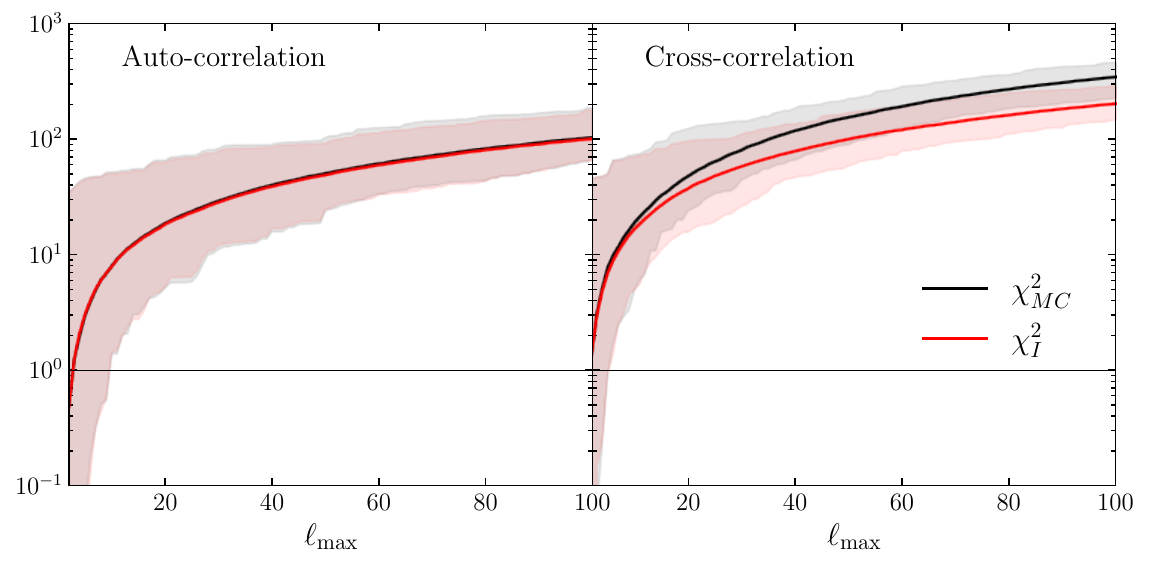}
  \captionsetup{width=\textwidth}
  \caption{\justifying Cross-correlations enable us to do model comparison. Here we show distributions of intrinsic $\chi^2_{I,r}$ and model-comparison $\chi^2_{\rm MC,r}$ for all realizations. Shaded regions show the full distribution bounds, with medians highlighted. These distributions are used to compute the D coefficients in Figure~\ref{fig:auto_cross}.
  \textit{Left panel:} The auto-correlation case  shows overlapping distributions, indicating no significance when comparing a realization to a fiducial model. \textit{Right panel:} In the cross-correlation setting, $\chi^2_{\rm MC,r}$ exceeds $\chi^2_{I,r}$ when cross-correlations and small angular scales are included, enabling significant model comparison. } 
  \label{fig:bottom}
\end{figure}

In figure~\ref{fig:bottom}, we illustrate the intrinsic~$\chi^2_{I,r}$ and model-comparison~$\chi^2_{\rm MC,r}$ distribution for all our realizations. 
These distributions are used to compute the D coefficients shown in figure~\ref{fig:auto_cross}.
We highlight the median of each quantity, and the
shaded region represents the full bounds of the distribution across realizations.
The~$\chi^2_{I,r}$ coefficient quantifies how much a realization of a model is statistically distinguishable from the average realization of a fiducial model. 
This is compared with the intrinsic scatter of the realizations of a given theory (encoded in $\chi^2_{I,r}$) to build up the D statistics from equation~\eqref{eq:Ddef}.
In the left panel, the two distributions are fully compatible; hence we cannot distinguish a realization of the fiducial model and one from a different model. 
On the other hand, when including cross-correlations and sufficiently small angular scales, $\chi^2_{\rm MC,r}$ is consistently above the~$\chi^2_{I,r}$ distribution, implying that the model comparison can be carried out with high significance. 

%%%%%%%%%%%%%%%%%%%%%%%%%%%%%%%%%%%%%%%%%%%%%%%%%%%%%%%%%%%%%%%%%%%%%%%%%%%%%%%%%%%%%%%%%%%%%%%%%%%%%%%%%

\subsection{Equivalent chi-squared formulation}

It is well-known that matrix properties allow for an alternative mathematical formulation of the~$\chi^2_{\mathrm{MC},r}$ appearing in equation~\eqref{eq:snrAB}, see, e.g., the detailed derivation of ref.~\cite{Hamimeche_2008}.
In practice, $\chi^2_{\mathrm{MC},r}$ can also be expressed as
\begin{equation}
    \chi^2_{\mathrm{MC},r} = f_{\mathrm{sky}}\sum_2^{\ell_\mathrm{max}}  (\Delta\mathbf{\Sigma}_\ell)^T \mathcal{M}^{-1}_\ell (\Delta\mathbf{\Sigma}_\ell)\,,
\label{eq:delta_chi2}
\end{equation}
where~$\Delta\mathbf{\Sigma}_\ell = \mathbf{C}^\mathrm{(r,A)}_\ell-\mathbf{C}^{\rm (th,B)}_\ell$, each 1D vector~$\mathbf{C}_\ell$ contains the same information as the covariance matrix~$\mathcal{C}_{\ell}$, and it is organized as
\begin{equation}
    \mathbf{C}_\ell = 
    \left( \begin{matrix}
    C_\ell^{\mathrm{gal},\mathrm{gal}}(z_1,z_1) \\
    \vdots	\\
    C_\ell^{\mathrm{gal},\mathrm{GW}}(z_1,f_1) \\
    \vdots	\\
    C_\ell^{\mathrm{GW},\mathrm{GW}}(f_1,f_1) \\
    \vdots
    \end{matrix} \right)\,.
\label{eq:1d_cl}
\end{equation}
Due to the symmetry of $\mathcal{C}_\ell$, only the diagonal and upper-diagonal entries are included. 
In order to be properly combined with the data vector of equation~\eqref{eq:1d_cl}, the new covariance matrix $\mathcal{M}_\ell$ is constructed by associating to every index $I$ and $J$ of the column vector $\mathbf{C}_\ell$ a couple of indeces ($I_1$, $I_2$) and ($J_1$, $J_2$), to encode the two tracers in a given redshift or frequency, e.g.,~for $C_\ell^{\rm {\rm gal,GW}}(z_1,f_1)$, $I_1={\rm gal}_{z_1}$ and $I_2={\rm GW}_{f_1}$. 
The new covariance matrix is then defined as
\begin{equation}
    \label{eq:Mlijcov}
    \mathcal{M}_\ell^{IJ}\!=\!\frac{1}{2 \ell+1}\qty(C_{\ell}^{\qty(I_1, J_1)} C_{\ell}^{\qty(I_2, J_2)}\!+\!C_{\ell}^{\qty(I_1, J_2)} C_{\ell}^{\qty(I_2, J_1)}),
\end{equation}
which can be generically estimated in a simulation-based analysis as~\cite{Giannantonio_2008,Ho_2008}
\begin{equation}
  \widehat{\mathcal{M}}_\ell^{IJ} = {\rm Cov} \left( C_\ell^{I_1I_2}, C_\ell^{J_1J_2} \right) = \frac{1}{N_r} \sum_{k = 1}^{N_r} \left[ \hat{C}_{\ell,k}^{I_1I_2} - \bar{C}_\ell^{I_1I_2} \right] \left[ \hat{C}_{\ell,k}^{J_1J_2}  - \bar{C}_\ell^{J_1J_2}  \right],
\end{equation}
where the bar denotes the ensemble average and the hat denotes the realization-dependent estimate.
Due to the statistical properties of the spherical harmonic coefficients, it immediately follows that~$\left\langle \widehat{\mathcal{M}}_\ell^{IJ} \right\rangle = \mathcal{M}_\ell^{IJ}$.

%%%%%%%%%%%%%%%%%%%%%%%%%%%%%%%%%%%%%%%%%%%%%%%%%%%%%%%%%%%%%%%%%%%%%%%%%%%%%%%%%%%%%%%%%%%%%%%%%%%%%%%%%

\section{Kolmogorov-Smirnov test of angular power spectra distribution}
\label{app:ks_test}

We report the results of the Kolmogorov-Smirnov tests used to check whether the GWB angular power spectra follow a Wishart distribution for different source-removal thresholds and multipoles.
We compare the distribution of~$\hat{C}_\ell$ from our simulated GWB realizations with the theoretical Wishart distribution expected for Gaussian fields, for different values of~$\ell$ and~$h_{\rm res}$.
For each case, we compute a $p_\ell$-value, which quantifies the probability that the observed distribution of~$\hat{C}_\ell$ is consistent with the Wishart distribution at that multipole for that choice of~$h_{\rm res}$.

\begin{figure}[ht]
    \centering
    \includegraphics[width=0.8\textwidth]{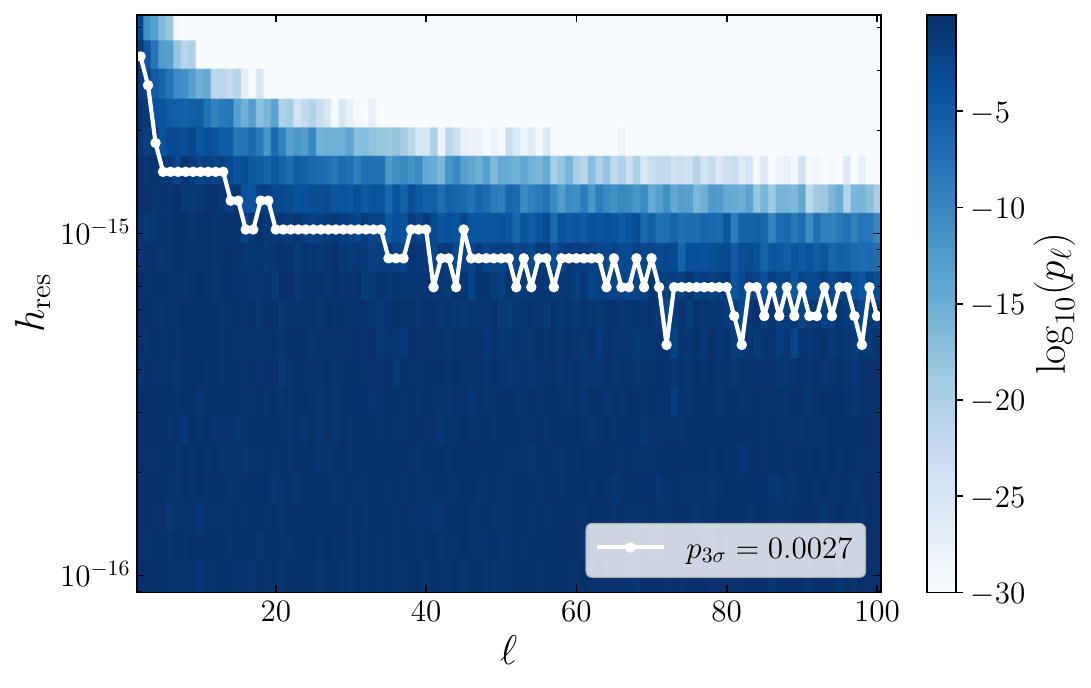}
    \caption{Kolmogorov-Smirnov test of the GWB $C_\ell$ distribution. 
    The heatmap shows the~$p$-values of the test comparing the simulated~$\hat{C}_\ell$ distribution with the theoretical Wishart distribution, as a function of source-removal threshold~$h_{\rm res}$ and multipole~$\ell$. 
    The white line indicates the boundary where~$p_\ell = p_{3\sigma}$, defining the maximum~$h_{\rm res}^{\max}(\ell)$ still compatible with a Wishart statistical distribution at the~$3\sigma$ confidence level.}
    \label{fig:ks_test_heatmap}
\end{figure}

In figure~\ref{fig:ks_test_heatmap} we show the resulting $p_\ell$-values of the test and we highlight for each multipole at which sensitivity the $p$-value exceeds the~$3\sigma$ threshold, i.e., when~$p_\ell\geq p_{3\sigma}=0.0027$.
In that region, the distribution of angular power spectra is indistinguishable from a Wishart.
As expected, small angular scales (high $\ell$) require more stringent source-removal thresholds to maintain the Gaussianity of the GWB field, since the variance of the $\hat{C}_\ell$ distribution is tighter at high $\ell$ than at low multipoles.
For the largest multipole~$\ell_{\rm max}=70$ used in the analysis, we find that~$h_{\rm res}\lesssim 8.6\times 10^{-16}$ is required to ensure Wishart-distributed GWB statistics at~$3\sigma$ confidence.
This threshold value also ensures Gaussianity for all multipoles~$\ell \leq\ell_{\rm max}$, and is well above the~$h_{\rm res}=10^{-16}$ conservative threshold value adopted in this work.

%%%%%%%%%%%%%%%%%%%%%%%%%%%%%%%%%%%%%%%%%%%%%%%%%%%%%%%%%%%%%%%%%%%%%%%%%%%%%%%%%%%%%%%%%%%%%%%%%%%%%%%%%

\section{Impact of GWB bias on detectability}
\label{app:gwb_bias_detectability}

The statistical significance of the~$D$ metric is inherently connected with the clustering strength of both tracers included in the analyses.
The galaxy population used in this work has quite standard clustering properties, as discussed in appendix~\ref{app:class_validation}.
On the other hand, the GWB bias still represents an unknown property; thus, in this appendix we present an estimate of the detectability coefficient dependence on GWB bias.

Since the methodology of this paper is entirely simulation-based, varying the bias of the GWB would mean at the practical level exploring the parameter space underlying the HOD, \textsc{UniverseMachine}, and SMBHB models.
The number of parameters is at least~$\mathcal{O}(40)$; and each point of such parameter space would require running multiple realizations, as done in this work, making the task computationally prohibitive.
Therefore, here we resort to an alternative methodology, which allows us to create synthetic angular power spectra and maps, with tunable values of galaxy and GWB biases.

First, we derive from our set of simulations the average angular power spectra~$\bar{C}_\ell^{\rm gal,gal}$, $\bar{C}_\ell^{\rm GW,GW}$, and~$\bar{C}_\ell^{\rm gal,GW}$ in the weak-field limit ($h_{\rm res}=10^{-16}$).
These mean spectra are further smoothed to remove residual fluctuations.
Following ref.~\cite{Giannantonio_2008}, we then build synthetic maps by sampling harmonic coefficients for each~$(\ell\geq 2, m\geq 0)$ as
\begin{equation}
    a_{\ell m}^\mathrm{gal} = \xi_a \sqrt{\bar{C}_\ell^\mathrm{gal,gal}}\,, \qquad
    a_{\ell m}^\mathrm{GW} = \xi_a \frac{\bar{C}_\ell^\mathrm{gal,GW}}{\sqrt{\bar{C}_\ell^\mathrm{gal,gal}}} + \xi_b \sqrt{\bar{C}_\ell^\mathrm{GW,GW} - \frac{(\bar{C}_\ell^\mathrm{gal,GW})^2}{\bar{C}_\ell^\mathrm{gal,gal}}}\,,
\end{equation}
where~$\xi_a, \xi_b$ are uncorrelated unit-variance random variables.
We repeat this process to create 1,000 synthetic maps for the fiducial angular power spectra obtained from simulations.
Then, we perform the same sampling for two additional scenarios, where the cross power spectra are~$\bar{C}_\ell^{\rm gal,GW}/2$ and~$2\bar{C}_\ell^{\rm gal,GW}$.
These scenarios correspond to the cases in which~$b_\mathrm{GW}=b^\mathrm{sim}_\mathrm{GW}/2$ and~$b_\mathrm{GW}=2b^\mathrm{sim}_\mathrm{GW}$, respectively, where~$b^\mathrm{sim}_\mathrm{GW}$ is the GWB linear bias obtained from simulations.
In the latter two scenarios, $\bar{C}_\ell^{\rm GW,GW}$ is left fixed to the simulation value, since it is completely shot noise dominated, thus a change in the bias of the clustering signal would produce a negligible change in the total angular power spectrum.

\begin{figure}[ht]
    \centering
    \includegraphics[width=0.6\textwidth]{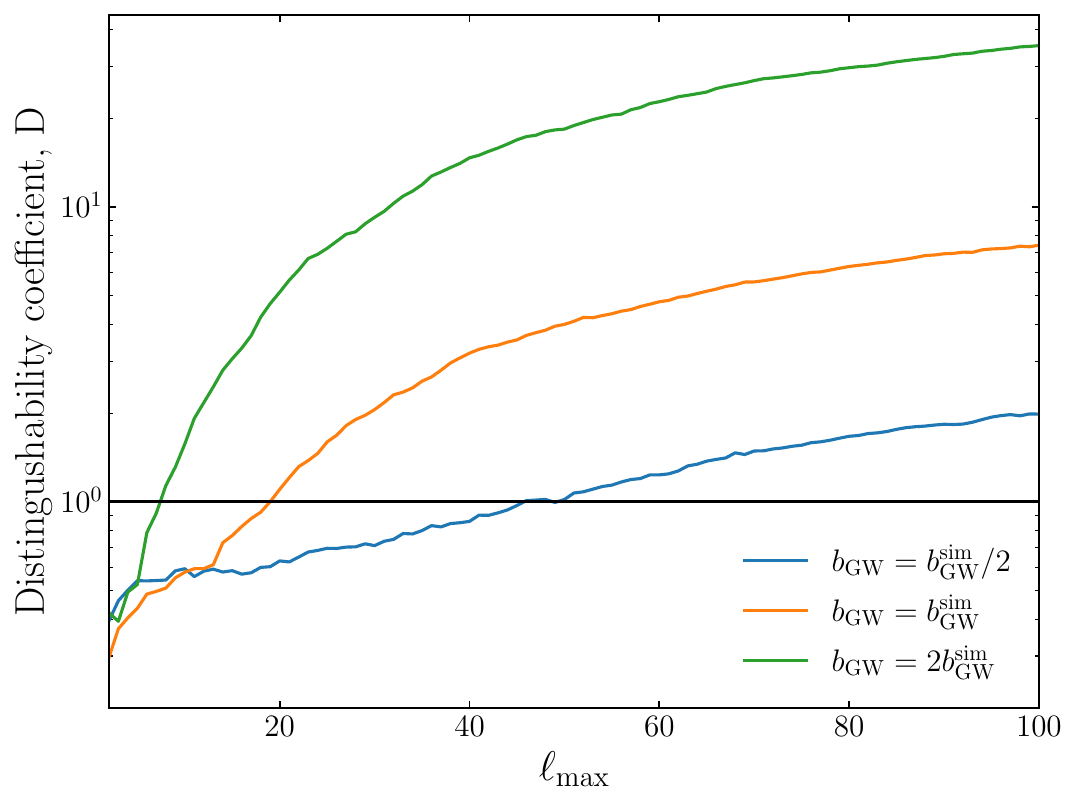}
    \caption{Median distinguishability coefficient $D$ (over 1000 GWB realizations) as a function of $\ell_{\rm max}$, for a representative set of rescaled GWB biases. 
    The test is performed in the weak-field regime ($h_{\rm res}=10^{-16}$) to ensure Wishart-distributed GWB statistics, and only the cross-correlation spectrum is varied to isolate bias dependence. 
    The results show a clear monotonic increase of $D$ with $f$, confirming that higher GWB bias leads to stronger cross-correlation signals and thus higher distinguishability.}
    \label{fig:bias_dependence}
\end{figure}

Finally, we run these synthetics realizations through our pipeline.
We show in figure~\ref{fig:bias_dependence} the median distinguishability coefficients for each scenario.
As expected, the median distinguishability for the fiducial scenario matches that reported in figure~\ref{fig:auto_cross}.
This results confirms the sensitivity of our statistical tool to the GWB bias, and showcases the importance of understanding how strongly SMBHBs are clustered.

%%%%%%%%%%%%%%%%%%%%%%%%%%%%%%%%%%%%%%%%%%%%%%%%%%%%%%%%%%%%%%%%%%%%%%%%%%%%%%%%%%%%%%%%%%%%%%%%%%%%%%%%%

\bibliography{apssamp}
\bibliographystyle{utcaps}

\end{document}